\documentclass[%
reprint,
superscriptaddress,
amsmath,amssymb,
 aps, prx,
floatfix
]{revtex4-2}

\usepackage{graphicx}
\usepackage{dcolumn}
\usepackage{bm}

\usepackage[inkscapelatex=false]{svg}
\usepackage{amsmath}
\usepackage[export]{adjustbox}
\usepackage{colortbl}

\usepackage{placeins} 

\newcommand{\figref}[1]{Fig.~\ref{#1}}

\newcommand{\secref}[1]{Section~\ref{#1}}
\newcommand{\tabref}[1]{Table~\ref{#1}}
\newcommand{\eqnref}[1]{Eq.~\eqref{#1}}
\newcommand{\appref}[1]{Appendix~\ref{#1}}

\begin{document}

\preprint{APS/123-QED}

\title{\textbf{Reconfigurable Superconducting Quantum Circuits Enabled by Micro-Scale Liquid-Metal Interconnects} 
}%

\author{Zhancheng~Yao}

\email{yao1996@bu.edu}

\altaffiliation[\\Present address: ]{
Quantum Science Center of Guangdong-Hong Kong-Macao Greater Bay Area, Shenzhen 518045, China
}

\affiliation{
Division of Materials Science and Engineering, Boston University, Boston, Massachusetts 02215, USA
}

\author{Nicholas~E.~Fuhr}
\affiliation{
Department of Electrical and Computer Engineering, Boston University, Boston, Massachusetts 02215, USA
}

\author{Nicholas~Russo}

\affiliation{ 
Department of Physics, Boston University, Boston, Massachusetts 02215, USA
}%
 
\author{David~W.~Abraham}

\affiliation{
IBM Quantum, IBM T.J. Watson Research Center, Yorktown Heights, New York 10598, USA
}%

\author{Kevin~E.~Smith}

\affiliation{ 
Division of Materials Science and Engineering, Boston University, Boston, Massachusetts 02215, USA
}

\affiliation{ 
Department of Physics, Boston University, Boston, Massachusetts 02215, USA
}%

\affiliation{ 
Department of Chemistry, Boston University, Boston, Massachusetts 02215, USA
}

\author{David~J.~Bishop}
\affiliation{ 
Division of Materials Science and Engineering, Boston University, Boston, Massachusetts 02215, USA
}

\affiliation{
Department of Electrical and Computer Engineering, Boston University, Boston, Massachusetts 02215, USA
}

\affiliation{ 
Department of Physics, Boston University, Boston, Massachusetts 02215, USA
}%

\affiliation{ 
Department of Mechanical Engineering, Boston University, Boston, Massachusetts 02215, USA
}%

\affiliation{ 
Department of Biomedical Engineering, Boston University, Boston, Massachusetts 02215, USA
}%


\date{\today}

\begin{abstract}
Modular architectures are a promising route toward scalable superconducting quantum processors, but finite fabrication yield and the lack of high quality temporary interconnects impose fundamental limitations on system size. Here, we demonstrate chip-scale liquid-metal interconnects that show promise for plug-and-play superconducting quantum circuits by enabling non-destructive module replacement while maintaining high microwave performance. Using gallium-based liquid metals, we realize high-quality inter-module signal and ground interconnects, comparable in performance to conventional coplanar waveguide resonators. We illustrate consistent device characteristics across three thermal cycles between room temperature and 15 mK, as well as the ability to reform superconducting connections following module replacement. A width-dependent resonance frequency shift reveals a significant kinetic inductance fraction, which we attribute to the presence of $\beta$-phase tantalum as confirmed by X-ray characterization. Finally, we investigate power-dependent loss mechanisms and observe high-power dissipative nonlinearities qualitatively consistent with a readout-power heating model. These results establish liquid metals as viable chip-scale interconnects for reconfigurable, modular superconducting quantum systems.
\end{abstract}

\maketitle


\section{\label{sec:intro}Introduction}

Superconducting quantum circuits are among the most promising platforms for realizing fault-tolerant quantum computation, as demonstrated by recent progress in the preparation of large entangled states \cite{javadi2025big} and the implementation of quantum error correction \cite{google2025quantum}. Scaling these systems to practical sizes, however, remains a major engineering challenge. Modular architectures offer a powerful strategy to address this challenge, as different levels of modularity can mitigate distinct bottlenecks in fabrication, packaging, and control \cite{bravyi2022}. In particular, chip-scale modularity provides an effective route to extending the size of quantum processors while preserving high-fidelity quantum operations.

Chip-scale modular approaches, including flip-chip integration, have been widely explored and experimentally demonstrated \cite{gold2021entanglement,conner2021superconducting,kosen2022building,zhao2022tunable,yost2020,das2018cryogenic,rosenberg2017,wu2024modular,field2024modular}. Despite these successes, finite fabrication yield and limited reproducibility impose a hard ceiling on the achievable system size when permanent interconnects are used. A natural path to overcoming this limitation is to enable \textit{plug-and-play} modularity, in which defective modules can be replaced without discarding the entire system. While several proposed architectures can potentially offer plug-and-play functionality, they typically rely on bulky packaging and long-range multimode cables \cite{niu2023low,qiu2025deterministic,mollenhauer2025high,zhou2023realizing}. These approaches require substantial cryogenic space, introduce additional thermal load, and risk parasitic coupling to adjacent cable modes, all of which may ultimately limit scalability.

An appealing alternative is to employ chip-scale, short-range interconnects that preserve plug-and-play capability while avoiding the drawbacks of long-distance wiring. Such interconnects can operate in the single-mode regime, thereby maintaining gate speed and fidelity, and can be implemented with minimal packaging overhead. Achieving these goals, however, requires interconnect technologies that are simultaneously low-loss, reusable, and compatible with cryogenic operation.

Liquid metals (LMs), which are liquid at room temperature, have been extensively studied in the context of wearable and stretchable electronics \cite{dickey2017stretchable}. Gallium-based alloys, such as EGaIn (78.6\% Ga and 21.4\% In by weight) and galinstan (68.5\% Ga, 21.5\% In, and 10\% Sn by weight), are particularly attractive due to their low toxicity and negligible vapor pressure compared with mercury \cite{dickey2017stretchable}. Recently, high-quality liquid-metal-bridged coplanar waveguide resonators (CPWRs) have been demonstrated, establishing LMs as viable alternatives to solid galvanic interconnects in superconducting quantum circuits \cite{yao2024low}. Moreover, the intrinsic fluid properties of LMs, including self-healing \cite{li2016galinstan} and self-aligning behavior \cite{ozutemiz2018egain}, make them especially well suited for reconfigurable and replaceable interconnects.

\begin{figure}[htb]
\centering
\includegraphics[width = \linewidth]{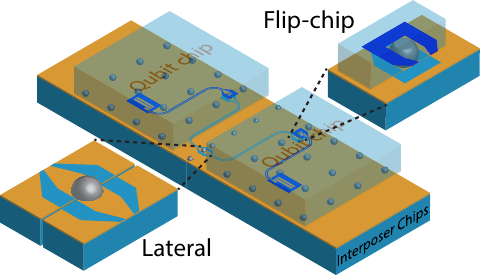}
\caption{\label{fig:sch} Example of a modular liquid-metal-enabled quantum system. The qubit chips and interposer chips are connected by superconducting liquid-metal bumps (gray). The adjacent half-bus resonators (cyan meandered line) on the interposer chips are bridged laterally by liquid metal at the chip edges. The qubit chips are self-spaced, self-aligned, and connected/grounded by liquid metal droplets in a flip-chip architecture. The qubit bus resonators (blue meandered line) are capacitively coupled to the qubits (blue parallel bars) and connected to the interposer by liquid metal on the other end. Non-destructive replacement of modules at room temperature is enabled by the liquid nature of the interconnects.
}
\end{figure}

These properties motivate a modular quantum architecture, illustrated schematically in Fig.~\ref{fig:sch}, in which chip-scale liquid-metal interconnects enable non-destructive module replacement at room temperature. After system-level testing and warm-up from the milliKelvin regime, defective modules could be removed, leaving residual liquid metal on the contact pads. New modules could then be installed by rebonding to the residual LM, allowing the system to be returned to operation at cryogenic temperatures. This approach provides a promising pathway toward scalable quantum processors without requiring disruptive advances in fabrication technology.

In this work, we demonstrate high-quality liquid-metal interconnects that function as both signal and ground interconnects across modular superconducting circuits, with performance comparable to conventional CPWRs. We further investigate the reusability of these interconnects, showing consistent device performance over multiple thermal cycles and the ability to form superconducting connections before and after module replacement. To confirm the fundamental mode of interconnected resonators, we design and observe a width-dependent resonance frequency shift indicative of a significant kinetic inductance fraction, which we attribute to the presence of $\beta$-Ta based on X-ray characterization. Finally, to elucidate the origin of the high fraction of power-independent loss, we probe high-power dissipative nonlinearities and find qualitative agreement with a readout-power heating model.

\section{\label{sec:modular}Liquid metal interconnects across modules}

\subsection{\label{sec:des}Chip assembly design}

\begin{figure*}
\includegraphics[width = \linewidth]{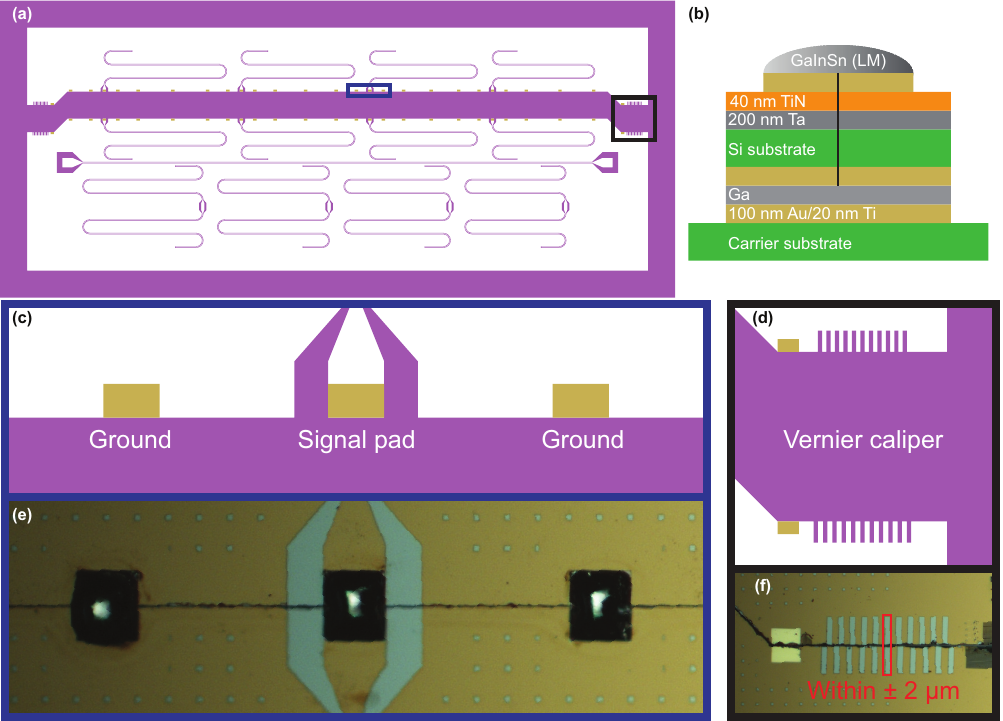}
  \caption{The chip assembly design: The top chip layout and material stack of the chip assembly are shown in (a) and (b), respectively. In the layout, the smaller chip at the top is chip A. It consists of four half-wavelength resonators, each extended from the matching edge. The larger chip at the bottom is chip B, which contains the other four halves of the resonator, all with matching edges and complementary shapes, allowing them to align with the four halves on chip A. The other four continuous resonators serve as control resonators. (c) Zoom-ins of the LM pads shown in (a). To prepare for the subsequent liquid metal deposition, 25-by-50~$\mathrm{\mu m}$ gold-on-titanium pads are defined at the end of where the center strips of half-resonators meet the matching edge. Identical ground pads are also defined with less than 500~$\mathrm{\mu m}$ pitch size. (d) Zoom-ins of the vernier scale markers shown in (a). The vernier scale markers at the edges of adjacent chiplets visualize the lateral alignment. (e) The LM droplets adhere to the gold pad shapes. (f) The alignment accuracy is within $\pm$2~$\mathrm{\mu m}$, visualized by the aligned central markers.}
  \label{fig:layout}
\end{figure*}

Our tests will be with resonators and use the quality factor as the metric for high performance. To test the LM performance across two modules, we used LM for both signal paths and ground interconnects. Here, we refer to the smaller module at top as chip A, and the larger module at bottom as chip B, shown in \figref{fig:layout}(a). Chip A consists of four halves of lithographically defined half-wavelength ($\lambda/2$) CPWRs. The other four halves are defined on chip B and later bridge to the ones on chip A using LM. Chip B also includes four $\lambda/2$ control resonators coupled to one common transmission line (TL). All the superconducting circuits were defined on a 200-nm tantalum layer with a 40-nm titanium nitride passivation layer as shown in \figref{fig:layout}(b). Two halves of LM pad array were patterned on top of Ta/TiN layers. These LM pads are 50-$\mathrm{\mu m}$ wide gold pads with a titanium seed layer and were patterned on the matching edge of the signal path and ground of the LM-bridged CPWRs as shown in \figref{fig:layout}(c). They define the LM interconnect shapes in the later selective LM deposition process as shown in \figref{fig:layout}(e). The distance between the signal LM bumps and their nearest ground LM interconnect affects the frequency of the LM-bridged resonators, so their distance was calibrated to match the designed resonator frequency.  An array of 5 $\mathrm{\mu m}$ vortex trapping holes with 25 $\mathrm{\mu m}$ pitch size is distributed across the ground plane. Plentiful wire bonds connect the ground conductors across the CPWRs and to the PCB ground, minimizing crosstalk and slotline modes. No wire bonds were placed across the matching edge in order to emulate the scenario when only LMs connect two chips.

Twelve chip assemblies, comprising six short-ended and six open-ended resonators, were fabricated to investigate different loss mechanisms. All the resonators and TLs have center strip widths and gap sizes of 10 and 6 $\rm{\mu m}$, respectively, except for the wire-bonding pads at the two ends of the TL and LM pads on the matching edges. Each pair of control and LM-joined resonators have resonance frequencies spaced by about 100 MHz, although it has been shown that the fundamental modes of these resonators are dropped by about half due to the large kinetic inductance fraction. This will be discussed in \secref{sec:KI}. Their coupling quality factor to the common TL is on the order of $10^5$, which is designed to be close to their internal quality factor for better fitting.

Each side of the matching edge of chips A and B has a complementary shape to the other, facilitating alignment during chip assembly. A carrier chip is on the bottom of chips A and B. We used gallium as the adhesive between the top chips and their carrier chip. The gallium, sandwiched between the top and carrier chips in a approximately 30~\textdegree C diluted hydrochloric acid (HCl) solution, turned to its liquid form without perturbation of its oxide. In this form, the gallium possesses a self-aligning property to also assist in the alignment process \cite{ozutemiz2018egain}. To help the gallium adhere firmly to the top and carrier chips in the HCl solution, the back side and top chips, as well as the top side of the carrier chip, are coated with Au/Ti. Liquid gallium adheres more readily to the gold surface than to silicon. The gold pad on the carrier chip should have a similar shape to the assembled top chip for easier alignment \cite{mastrangeli2017surface}. Vernier scale markers (\figref{fig:layout}(d)) were patterned on both sides of the matching edge to visualize the alignment in the lateral direction. Once the alignment was done, the gallium was solidified by cooling it with liquid nitrogen and avoiding a metastable supercooled liquid gallium phase. If we would like to replace one of the chips, we can elevate the chip temperature slightly above the melting point of gallium and replace the unwanted chip with a new one. We chose gallium for its simplicity and compatibility with our process, but one could adjust the ratio of gallium to other metals (e.g., indium or tin) to modulate the melting point, thereby adapting their subsequent process temperature and thermal budget.

We have achieved an alignment accuracy within $\pm$2~$\mathrm{\mu m}$ in the lateral direction, quantified by the vernier scale markers shown in \figref{fig:layout}(f),  facilitated by the self-aligning property without any precise pick-and-place tools. If the central markers align, the alignment accuracy is within $\pm$2~$\mathrm{\mu m}$. If the alignment shifts to one marker to the right, the alignment accuracy is estimated to be between $+$2 and $+$4~$\mathrm{\mu m}$; one marker to the left indicates an alignment accuracy of $-$2 to $-$4~$\mathrm{\mu m}$, and so on. These experiments required several rounds of trial and error to achieve the desired accuracy in alignment. The reason is that we do not examine chips under the microscope until they are free from HCl solution and dried with nitrogen. Moreover, adhesion and friction exert a strong holding force against self-alignment, especially when the gallium layer is sufficiently thin. However, a thin layer is still favored as excess gallium could protrude and then force a capillary-driven separation of the previously flush top chips. Thus, we traded the smooth self-align process off for a more flush top chip surface. Nevertheless, achieving this micron-level precision without the aid of a delicate tool demonstrates the potential of combining the matching shapes and self-aligning properties of liquid gallium. We chose this method because we want to allow for plug-and-play modularity; however, the space constraints of our PCB limit the available options. This method requires no additional fixture structure in the lateral dimensions and thus works well in limited space. 

\subsection{\label{sec:LMinter}Liquid metal interconnects loss in low-power regimes}

\begin{figure}[tb!]
\centering
\includegraphics[width = \linewidth]{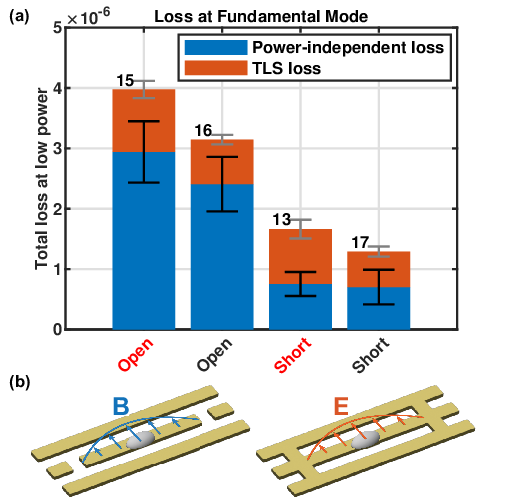}
  \caption{The loss of LM and control resonators for open- and short-ended resonators. (a) The total loss at low power, 15~mK. The heights of the bars denote the sample mean of the loss. The colors of the bars depict the type of loss, where the orange bars are TLS loss and the blue bars are power-independent loss. The $x$-axis labels read the type of the resonators, either open- or short-ended resonators, and the colors of the labels represent whether the samples are LM (red) or control (black) resonators. The error bars centered at the top of each stacked bar are the standard errors of the sample mean. The black error bars are for power-independent loss, and the grey error bars are for TLS loss. The number of samples for each type is labeled above the corresponding bar. (b) The schematic of open- and short-ended resonators with the LM droplets at $\lambda/4$. The lefthand cartoon is the open-ended resonators, where the $\lambda/4$ section exhibits the strongest magnetic field at the fundamental mode. The righthand one is the short-ended resonators, where the $\lambda/4$ section exhibits the strongest electric field.}
  \label{fig:loss}
\end{figure}

To assess the practicality of the envisioned system in \figref{fig:sch}, we aim to determine the impact on performance upon introducing LM. We first compare the total loss of LM-bridged resonators across modules and continuous control resonators on the same module at low power, measured at 15~mK. We used the same set temperature for all of the microwave characterization in this work. The model we used to extract the two-level system (TLS) and power-independent quality factors in the linear regime is
\begin{eqnarray}
\frac{1}{Q_i}=\frac{1}{Q_\mathrm{TLS}}+\frac{1}{Q_\mathrm{other}},
\label{eqn:Qi}
\end{eqnarray}
where $Q_i$ is the internal quality factor, $Q_\mathrm{TLS}$ and $Q_\mathrm{other}$ are the TLS and power-indepedent quality factors, respectively. We only fit the data using \eqnref{eqn:Qi} in the linear regime. The power-independent quality factor is treated as a fit parameter, whereas the TLS quality factor is written as \cite{mcrae2020materials,chiaro2016dielectric}
\begin{eqnarray}
Q_\mathrm{TLS} = Q_\mathrm{TLS,0}\frac{\sqrt{1+\left(\frac{\langle n\rangle}{n_c}\right)^{\alpha_\mathrm{TLS}}}}{\operatorname{tanh}\left(\frac{hf_{r0}}{2k_BT}\right)},
\label{eqn:QTLS}
\end{eqnarray}
where $Q_\mathrm{TLS,0}$ is the inverse intrinsic TLS loss, $\langle n\rangle$ is the average photon number, $n_c$ is the characteristic photon number of the TLS saturation, $\alpha_\mathrm{TLS}$ is a coefficient accounting for a slower power dependence of the TLS saturation, $f_{r0}$ is the resonance frequency of the resonator in the linear regime, $h$ is the Planck constant, $k_B$ is the Boltzmann constant, and T is the temperature. Here, we assume the temperature is a constant of 15 mK, and the average photon number is given by
\begin{eqnarray}
\langle n\rangle = \frac{2}{\hbar\omega_{r0}^2}\frac{Z_0}{Z_r}\frac{Q_l^2}{Q_c}P_g,
\label{eqn:n_bar}
\end{eqnarray}
where $\hbar$ is the reduced Planck constant, $\omega_{r0}=2\pi f_{r0}$ is the angular resonance frequency of the resonator in the linear regime, $Z_0$ and $Z_r$ are the characteristic impedance of the TL and resonator, respectively, and are assumed to be 50 $\Omega$, $Q_l$ is the loaded quality factor, $Q_c$ is the coupling quality factor of the resonator to the TL, and $P_g$ is the input power at the TL, which is calculated from the source power given a $-75$~dB attenuation. It is convenient to add different types of loss together for comparison, as shown in \figref{fig:loss}(a), so we took the inverse of all the quality factors as the loss, $\delta \approx \operatorname{tan}\delta = 1/Q$. Here, the power-independent loss is the inverse of the $Q_\mathrm{other}$, and the TLS loss at low power is assumed to be $1/Q_\mathrm{TLS,0}$, as both the numerator and the denominator in \eqnref{eqn:QTLS} are approximately one when $T=15$~mK, $f_{r0}$ is 3-9~GHz, and $\langle n\rangle \rightarrow - \infty$. Some data are too noisy to reliably extract fit parameters, so we only keep the data points if the $95\%$ confidennce intervals of both TLS and power-independent quality factors do not include zero. The data in \figref{fig:loss}(a) were acquired in the third measurement, as we did not anticipate a significant drop in the resonance frequencies and thus only measured a narrower frequency range in the first two measurements. The reason of frequency drop will be discussed in \secref{sec:KI}.

As shown in \figref{fig:loss}(a), we observe that the sample mean of total LM resonator loss at low power is of the same order of magnitude as that of the total loss of the controls. For all types of resonators, the power-independent loss is at least comparable to TLS loss, even at sufficiently low power, unlike many previous studies \cite{gao2008,crowley2023disentangling,mcrae2020materials,muller2019towards}. For open-ended resonators, the power-independent loss seems to be even more dominant than TLS loss for both LM and control resonators. As we do not observe a significant increase in loss of LM resonators compared to controls, this significant power-independent loss is likely from the solid materials. The short-ended resonators are much less lossy, so the difference between the two types of resonators could shed light on the root cause. For open-ended resonators, the quarter-wavelength ($\lambda/4$) part, where the strongest magnetic field or current is exhibited, has a larger width compared to the short-ended ones, meaning that the inductive/resistive loss that increases with increasing width could be the dominant loss. On the other hand, the capacitive parts of the resonators for short-ended resonators are narrower, implying that the capacitive loss that decreases with increasing width could also be dominant. Future experiments are necessary to parameterize the sources of loss. Overall, we have not observed a significant degradation in performance when bridging two modules using LM interconnects.

\subsection{\label{sec:reuse}Reusability of liquid metal interconnects}

The next step to assess the practicality of the LM-enabled modular quantum computing processor is to determine the reusability of the LM interconnects. Whenever a replacement of the defective qubit modules is necessary, the entire processor must warm up to the room temperature and then cool down again once the replacement is complete. Thus, a consistent performance of the untouched LM interconnects over multiple thermal cycles, as well as a comparable performance of the replaced LM interconnects before and after module replacement, is preferred for the proposed modular architecture. Here, we demonstrate the reusability of the LM interconnects in the two mentioned aspects.

\subsubsection{\label{sec:aging}Liquid metal interconnects performance over multiple thermal cycles}

We measured the same samples three times over a two-month period to test the performance of LM interconnects in the low-power regime, where the first measurement was conducted two months after the samples were fabricated. No special treatment was applied to preserve the samples outside the dilution refrigerator. However, as mentioned earlier, we did not anticipate a significant drop in the resonance frequencies of the samples due to the large kinetic inductance fraction (see \secref{sec:KI}) when we first measured them, so not every resonator was measured until later. Therefore, we narrow down the range of frequencies to ensure that we compare the performance of the same set of resonators over thermal cycles.

\begin{figure}[htb!]
\includegraphics[width = \linewidth]{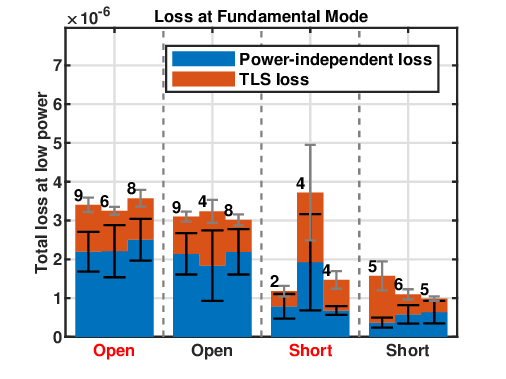}
\caption{\label{fig:aging} The total loss of LM and control resonators for open- and short-ended resonators at low power, 15~mK. The bar heights and colors denote the sample mean and type of loss, respectively, where blue bars represent power-independent loss and orange bars represent TLS loss. The number above each bar represents the number of samples. The lengths of the black and grey error bars indicate the standard errors of the sample means for the power-independent loss and TLS loss, respectively. The labels on the $x$ axis indicate the category of resonators, whether they are open- or short-ended, and whether they are LM (red) or control (black). Each category contains three bars, representing the first, second, and third measurements in chronological order from left to right. The first and third measurements are two months apart.
}
\end{figure}

As shown in \figref{fig:aging}, the LM and control resonators were comparable in each measurement, and they were relatively stable over time, except the short-ended LM resonators in the second measurement. The inconsistency may have been skewed by random fluctuations due to the small sample size. Nevertheless, we did not observe a significant aging over three thermal cycles and two months.

\subsubsection{\label{sec:sc}Resistance measurements before and after replacing modules}

In addition to the mentioned aging test, we also measured the resistance of the LM interconnect before and after replacing one of the modules.

\begin{figure}[htb!]
\includegraphics[width = \linewidth]{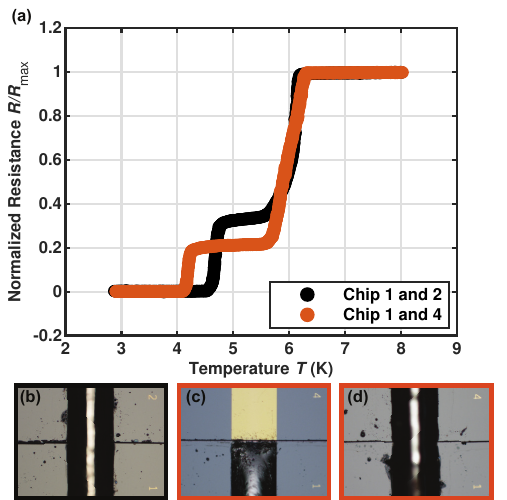}
  \caption{Resistance measurements before and after replacing modules. (a) The normalized resistance vs. temperature. The measurement starts at the base temperature of 3 K and gradually increases the temperature of the cold finger at a sufficiently low rate (1 mK/s). The colors represent whether the measurement is conducted before (black) or after (orange) the module replacement. (b)--(d) The photographs of the module replacement flow. (b) shows the LM interconnect between chips 1 and 2, and the resistance of this LM interconnect is indicated by the black data points in (a). (c) was taken after removing chip 2 and installing chip 4. The gold pads are used to define the LM shape. (d) Reconnecting the LM interconnect across chip 1 and the newly installed chip 4 (orange data points). The discrepancy of the resistance curve between the two measurements is likely due to the slight composition change in the LM interconnect induced by the HCl solution treatment for residue removal.}
  \label{fig:scba}
\end{figure}

We adopted a similar chip design as described in \secref{sec:des}, except we only used silicon chips and patterned gold pads for wirebond pads and LM adhesion. We first aligned the two top chips (chips 1 and 2) in a diluted HCl solution with gallium backside adhesive onto the carrier chip. Once the alignment is satisfactory, we fixed the alignment by freezing the assembly in liquid nitrogen. Then, we applied LM onto the predefined gold pads and created an interconnect across chips 1 and 2, as shown in the left photograph in \figref{fig:scba}(b). The chip assembly was wire-bonded to a PCB, and the LM interconnect resistance was measured using a four-point probe method at temperatures ranging from 3 to 8 K. The result is shown in the black-filled dots in \ref{fig:scba}(a). Furthermore, the chip assembly was warmed to room temperature and demounted from the PCB, then heated slightly above the melting point of gallium, and chip 2 was removed. We followed a similar alignment process to attach chip 4 adjacent to chip 1 as shown in \figref{fig:scba}(c). We applied more LM onto chip 4 and reconnected the LM interconnect as shown in \ref{fig:scba}(d), and conducted another four-point probe measurement. The result is shown in the orange-filled dots in \figref{fig:scba}(a).

We observed virtually zero resistance around 4 to 5 K for the LM interconnect both before and after module replacement, as shown in \figref{fig:scba}(a), despite a discrepancy between the two curves.
This discrepancy may be explained by composition differences induced by HCl etching, which can remove not only the surface oxide of gallium and its alloys but also the metals themselves. 
Further, HCl removes gallium and indium faster than tin \cite{kim2013recovery}, potentially altering the composition ratio. 
Since different ratios of gallium, indium, and tin can exhibit different superconducting transition curves \cite{ren2016}, such a shift could account for the observed variation.
Future experiments could be conducted to identify and confirm the root cause. 
Nevertheless, the LM interconnect reached virtually zero resistance before and after module replacement.
Combined with the aging test, this result demonstrates the preliminary reusability of the gallium-based LM interconnect.

\section{\label{sec:KI}Kinetic inductance related to $\beta$-tantalum}

We designed the resonators to operate at resonance frequencies of 6-8 GHz. However, our measurement algorithm detected and measured not only resonance responses close to the designed frequencies, but also the response at roughly half of the designed frequencies. 
We hypothesized two possible explanations as follows. 
First, since the lower and higher frequency responses are approximately half and equal to the designed frequencies, respectively, this could correspond to the superharmonic response of order two induced by forced and damped anharmonic oscillators with quadratic nonlinearity \cite{gusso2018approximate}. 
In this scenario, the resonators oscillate at twice the probe frequency when the probe frequency is near one-half of the natural frequency, corresponding to the responses at lower frequencies. Meanwhile, the responses near the designed frequencies are fundamental modes, as this nonlinearity should not significantly alter the natural frequencies.
As we observed nonlinearity in the resonance responses (see \secref{sec:hipwr}), this explanation is plausible if the quadratic term persists and remains strong even at low power.

The second explanation is that a significant fraction of the kinetic inductance of the resonator metalization induced a substantial decrease in the natural frequencies \cite{gao2008,porch2005calculation}, where the reduced resonance frequencies happen to be roughly half of the designed frequencies. Accordingly, the lower frequency responses are the fundamental modes, while the higher frequency responses near the designed frequencies are the first harmonics.
As we analyze the data based on the electrical and magnetic environments of LM droplets or solid materials, whether the responses correspond to fundamental modes will impact our analysis. 
For example, we assume the major contribution of TLS loss is from LM droplets for LM-bridged short-ended resonators since the electric field is the strongest near $\lambda/4$ at the fundamental mode. 
However, the TLS contribution should be from $\lambda/8$ and $3\lambda/8$ at its first harmonics, and LM droplets no longer play a crucial role as the electric field is almost zero near $\lambda/4$. 
Therefore, it is essential to understand the root cause of such low-frequency responses to interpret the data accurately.

To distinguish between the two hypotheses, we fabricated two sets of chips with resonators of the same length as those used in the LM low-power regime loss study, but with varying widths. To reduce the number of variables, we maintain the same ratio of center strip width to gap size between the center strip and the ground. If the first hypothesis (superharmonics) is accurate, we should see the same frequency distribution as before, since the length defines the resonance frequency. On the other hand, if the second explanation (kinetic inductance) prevails, the resonance frequency distribution should be different and depend on the center strip widths. The kinetic inductance fraction can be written as
\begin{eqnarray}
\alpha_L=\frac{L_k}{L}=\frac{gL_s}{L_m+gL_s}, g\propto 1/W,
\label{eqn:alpha_L}
\end{eqnarray}
where $L_k$, $L_m$, $L_s$, and $L$ are the kinetic, geometric, surface, and total inductance, respectively, $g$ is a geometry-dependent factor and is inversely proportional to the center strip width $W$ \cite{gao2008,porch2005calculation}. Thus, the calculated inverse of kinetic inductance fraction vs. center strip width should be linear if a significant kinetic fraction is the root cause. The experimental determination of $\alpha_L$ is relatively straightforward, assuming this is the correct physics model and $\alpha_L$ is large \cite{gao2008}:
\begin{eqnarray}
\alpha_L=1-\left(\frac{f_{r0}^{m}}{f_{r0}^{d}}\right)^2,
\label{eqn:alpha_Lexp}
\end{eqnarray}
where $f_{r0}^{m}$ and $f_{r0}^{d}$ are the measured and designed frequency, respectively.

\begin{figure}[htb!]
\includegraphics[width=\linewidth]{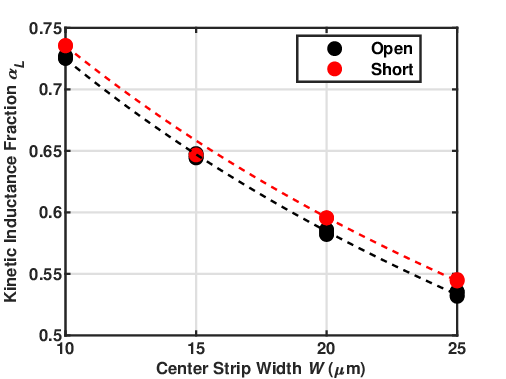}
\caption{\label{fig:KI} The calculated kinetic inductance fraction vs. center strip width. The black dots are from open-ended resonators, and the red dots are from short-ended resonators. The dashed lines with the corresponding color are the linear fit to $1/\alpha_L$ vs. $W$.
}
\end{figure}

We indeed observed width-dependent resonance frequencies, and the calculated $1/\alpha_L$ using \eqnref{eqn:alpha_Lexp} is found to be appreciably linear with respect to width. The inverse of fit results are shown in \figref{fig:KI} to depict $\alpha_L$ in the y-axis. Therefore, we attribute the low-frequency responses to the decrease in resonance frequencies resulting from a significant kinetic inductance fraction, and these responses are the fundamental modes of the resonators.

\begin{figure}[htb!]
\includegraphics[width=\linewidth]{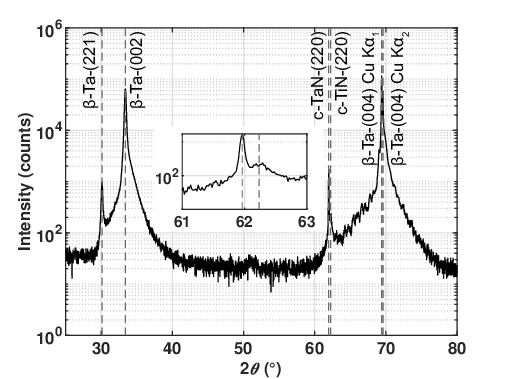}
\caption{\label{fig:XRD} X-ray diffractogram of 40 nm TiN/200 nm $\beta$-Ta/(001) Si substrate. The inset depicts the zoom-in peaks of c-TaN and c-TiN, both oriented in the (220) plane.
}
\end{figure}

To further understand the origin of the large kinetic inductance fraction, we performed X-ray diffraction of the TiN/Ta/Si, and the results are shown in \figref{fig:XRD}. 
We were able to identify the characteristic peaks of TiN and Ta. 
Specifically, the Ta peaks are from $\beta$-Ta.
We observed a peak significantly overlapping with the TiN peak, and we believe this peak was from the interfacial TaN.

\begin{figure}[htb!]
\includegraphics[width=0.9\linewidth]{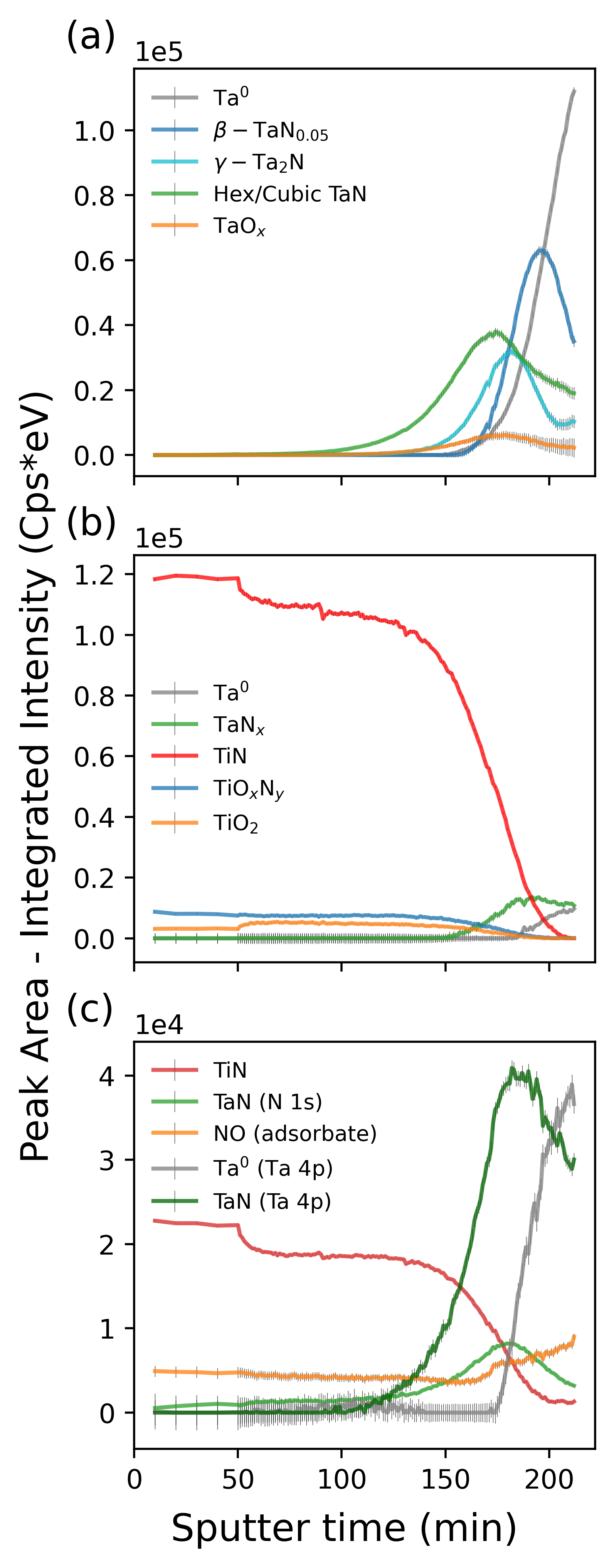}
\caption{\label{fig:XPSN2} Decomposition of XPS spectra for (a) Ta 4f, (b) Ti 2p, and (c) N 1s in terms of their component phases as a function of sputter time. A series of Ta phases appear in (a) before the Ta0 layer is reached. The Ti (b) and N (c) spectra overlap with Ta $\mathrm{4p_{1/2}}$ and Ta $\mathrm{4p_{3/2}}$, respectively, necessitating their inclusion for chemical analysis. The data in (b) and (c) show the near complete disappearance of Ti and N after 220 minutes of sputtering.
}
\end{figure}

These results were further supported by X-ray photoelectron spectroscopy (XPS) depth profile analysis, as shown in  \figref{fig:XPSN2}.
As we sputtered through the metalization layer, we started to see the TaN signature (green) when the ratio of TiN (gray) started to gradually decrease around 130 minutes of sputter time, as shown in \figref{fig:XPSN2}(c). 
This TaN phase appears before the TiN layer is completely removed, and before the Ta metal phase is reached there are significant changes in spectral structure and shifts to lower binding energies in the Ta 4f spectra, as shown in \figref{fig:XPSN1}(b).
We find that these spectral changes are well described by a model composed of TaN, $\gamma-$Ta$_2$N, $\beta-$TaN$_{0.05}$, and metallic Ta, which corresponds to preferential N sputtering and change in stoichiometry \cite{arranzAPACompositionTantalumNitride2005,arranzSIATantalumNitrideFormation2000}.

Although both XRD and XPS were used to characterize the  thin film material composition, we attribute the significant kinetic inductance fraction to the $\beta$-tantalum. First, we observed a frequency shift in both the LM and control resonators, so it is unlikely to be induced by the LM. Then, the metallization consists of a TiN encapsulation layer and a Ta base layer. In a previous study using a thinner TiN layer on an Nb base layer \cite{yao2024low}, we observed no significant shift, despite the thinner TiN likely having a higher kinetic inductance fraction.
Thus, the TiN encapsulation layer is unlikely to be the cause of the large fraction observed here. Therefore, the $\beta$-Ta base layer seems to be the variable. We cannot exclude the effect of the interfacial metallic compound, as they are thin layers; a thinner layer should possess a larger kinetic inductance fraction \cite{porch2005calculation}. 
This suggests a future control experiment with different thicknesses of the interfacial layer materials. However, if the significant kinetic inductance fraction is induced by $\beta$-tantalum, it is in agreement with the results reported by \cite{joshi2025investigating}.

\section{\label{sec:hipwr}Nonlinear resonance response in high-power regimes}

As discussed in \secref{sec:LMinter}, we observed that the power-independent loss in the system is at least comparable to the TLS loss in low-power regimes. Here, we discuss the high-power responses in an attempt to pinpoint possible loss channels.

We observed that both control and LM resonators exhibit strong nonlinear responses in the high-power regime. Therefore, the conventional diameter correction method \cite{khalil2012} fails, and we adopted the fitting procedures proposed in References \cite{swenson2013operation,dai2022new}. More details are discussed in \appref{sec:fitmod}.

\begin{figure}[ht!]
  \includegraphics[width=\linewidth]{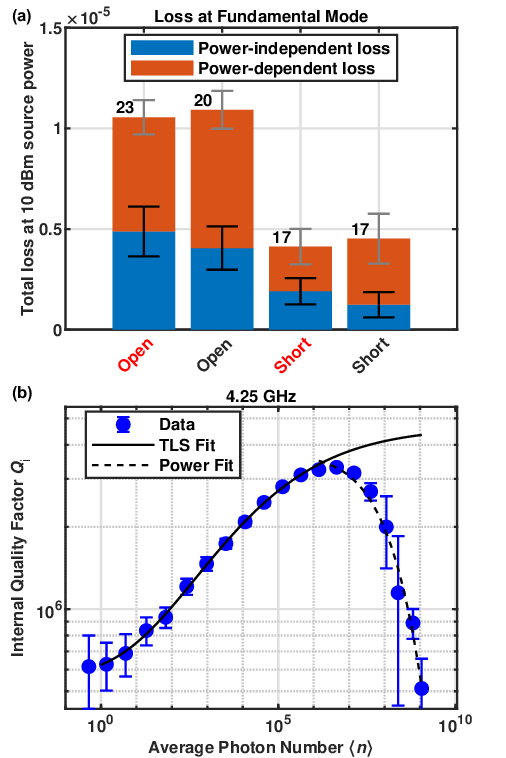}
  \caption{The loss of LM-containing and control resonators for open- and short-ended resonators in high-power regimes. (a) The total loss at a source power of 10~dBm at 15 mK. The heights of the bars denote the sample means of the loss. The colors of the bars depict the type of loss, where the orange bars are power-dependent loss and the blue bars are power-independent loss. The $x$-axis labels read the type of the resonators, either open- or short-ended resonators, and the colors of the labels represent whether the samples are LM (red) or control (black) resonators. The error bars centered at the top of each stacked bar are the standard errors of sample means. The thick, black error bars are for power-independent loss, and the thin, grey error bars are for power-dependent loss. The number of samples for each type is labeled above the corresponding bar. (b) The extraction of power-dependent and -independent loss. The point estimates of the internal quality factor, extracted from individual traces of resonance responses, are plotted against the average photon numbers as blue dots. The error bars represent the $95\%$ confidence intervals from the individual fits. The TLS model is implemented to fit the linear regime ($a_{n0} < 0.05$), and the fit result is depicted by the solid black curve. A simple power law function is used to fit the nonlinear data ($a_{n0} \geq 0.05$) in the log-log scale, and the fit result is depicted by the dashed black curve. The power-dependent loss at high powers, shown in (a), is calculated by the difference between the solid and dashed curves at 10~dBm.}
  \label{fig:losshi}
\end{figure}

We extracted the internal quality factor and corresponding $95\%$ confidence intervals, and plotted against the estimated average photon numbers. The internal quality factor decreases with increasing photon number (power) as shown in \figref{fig:losshi}(b) in high-power regimes. To quantify and compare such decreases among different types of resonators, we adopted the nonlinearity parameter, $a_{n0}$, to distinguish linear ($a_{n0} < 0.05$) and nonlinear ($a_{n0} \geq 0.05$) resonance responses and fit the two distinct regimes with different models. We adopted the conventional TLS model discussed in \secref{sec:LMinter} for linear regimes and a simple power function to fit the nonlinear data in the log-log scale, where $\operatorname{log}_{10}Q_i=k(\operatorname{log}_{10}\langle n \rangle)^b+c$, and $k$, $b$, and $c$ are fit parameters. We emphasize that this is \textit{not} a physics model but rather a phenomenological fit solely for comparing the quality factor or loss evaluated at common photon number or power. The power-dependent loss ($\delta_{\text{power}}$) in high-power regines shown in \figref{fig:losshi}(a) is calculated as
\begin{eqnarray}
\delta_\mathrm{power}=\frac{1}{Q_\mathrm{power}}=\frac{1}{Q_i}-\frac{1}{Q_\mathrm{other}}
\label{eqn:losspower}
\end{eqnarray}
at 10~dBm source power. We calculate the power-dependent loss with common power instead of photon number, since we did not find a common photon number overlap where every resonator is nonlinear. The open-ended resonators exhibit a more significant decrease in the internal quality factors compared to short-ended resonators. At 10~dBm source power, the power-dependent loss is more dominant than the power-independent loss. Again, we did not observe an obvious difference between the control and LM resonators. We suspect that the loss is due to readout-power heating \cite{de2010readout} and will discuss it later.

\begin{figure}[htb!]
    \includegraphics[width=\linewidth]{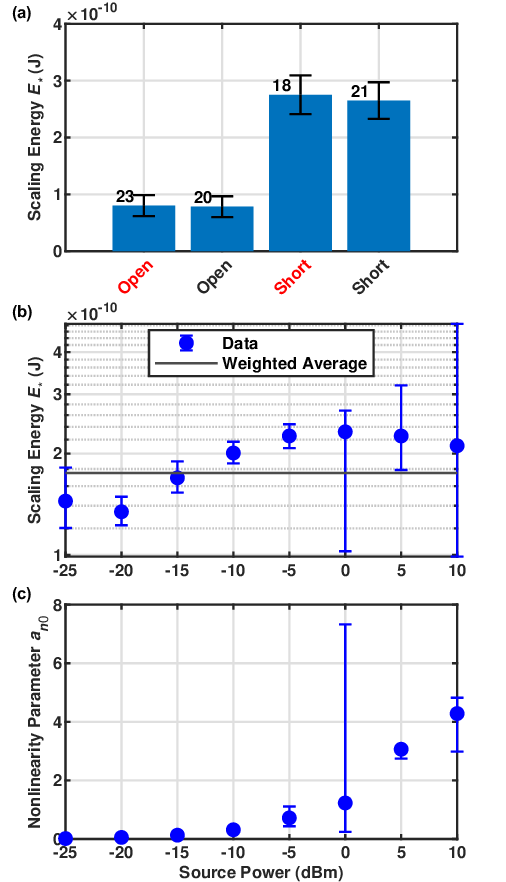}
    \caption{The scaling energy, $E_*$, and the nonlinearity parameter at $f_{r0}$, $a_{n0}$. (a) The sample means and standard errors of the sample means for different types of resonators. The $x$-axis labels read the type of the resonators, either open- or short-ended resonators, and the colors of the labels represent whether the samples are LM (red) or control (black) resonators. (b) The scaling energy extracted from linear fits to the stored energy of the resonator vs. the nonlinear fractional resonance frequency shift at different powers. The black line denotes the weighted average. (c) The nonlinearity parameters at different powers. The error bars in (b) and (c) are the $95\%$ confidence intervals derived from the bootstrap distribution.}
    \label{fig:Eas}
\end{figure}

As mentioned earlier, we distinguished between linear and nonlinear regimes using the nonlinearity parameter. Although $\beta$ in \eqnref{eqn:frb} is a measure of the nonlinearity, it is more convenient to compare material-dependent quantities like $E_*$ and $I_*$ \cite{shu2021nonlinearity}, and to quantify the extent of the bifurcation by the nonlinearity parameter
\begin{eqnarray}
a_n = \frac{2Q_l^3}{Q_c}\frac{P_g}{\omega_rE_*},
\label{eqn:an}
\end{eqnarray}
where $Q_l$ and $Q_c$ are the loaded and coupling quality factors, respectively, $\omega_r$ is the resonator circulating current-dependent angular resonance frequency, $E_*$ is the scaling energy dependent on materials and geometry. The inverse of $E_*$ measures the resonator nonlinearity given the same quality factor and applied power, so we compare the $E_*$ of each resonator directly. To extract the scaling energy, we performed a linear fit to the calculated stored energy in the resonator, $E$, vs. the nonlinear fractional resonance frequency shift, $\delta x$, and equated the scaling energy to the modulus of the slope. Then, the corresponding $a_{n0}$ ($a_n$ at $\omega_{r0}$) is derived. We resampled the fit parameters, such as $Q_l$ and $Q_c$, assuming a $t$-distribution derived from the fit, and recalculated $E_*$ and $a_{n0}$. We repeated the process $10^5$ times, and all the calculated values constitute the bootstrap distributions. The purpose of the bootstrapping process is to evaluate the confidence of the point estimates considering the uncertainty propagation from the previous fitting procedures and to use the confidence intervals to calculate the weighted average of $E_*$ shown in \figref{fig:Eas}(b). We treated the weighted average as the point estimate of each resonator and the inverse square of the confidence intervals as the weights. The point estimates and the $95\%$ confidence intervals of $a_{n0}$ at different powers are shown in \figref{fig:Eas}(c). The detailed procedures are discussed in \appref{sec:nlin}.

Further, we categorized each type of resonator and then calculated and compared the sample means and standard errors shown as the bar heights and error bars in \figref{fig:Eas}(a), respectively. Here, the open-ended resonators possess lower scaling energy than the short-ended ones, meaning the open-ended resonators exhibit stronger nonlinearity given the same power and quality factors. The scaling energy is expected to be on the order of the condensation energy of the superconductor if $\alpha_L \approx 1$ \cite{swenson2013operation}. The condensation energy of the superconductor is given by $E_\mathrm{cond}=N_0\Delta^2V/2$, where $N_0$ is the single spin density of the states at the Fermi energy, $\Delta \approx 3.5k_BT_c/2$ is the superconducting gap, and $V$ is the volume of the superconductor. If we equate $V$ to the volume of the center strip, $T_c=1$~K \cite{joshi2025investigating}, $N_0=2.678\times10^{10}$~$\mu \mathrm{m}^{-3}\mathrm{eV}^{-1}$ \cite{chi2024hybrid}, the coarse estimate of $E_\mathrm{cond}$ is about $9\times10^{-13}$~J, which is two to three orders of magnitude smaller than the sample means of the scaling energy. The decrease in the quality factor and orders of magnitude difference between $E_*$ and $E_\mathrm{cond}$ indicate that the nonlinearity is not purely reactive \cite{zmuidzinas2012superconducting}.

\begin{figure}[htb]
\includegraphics[width=\linewidth]{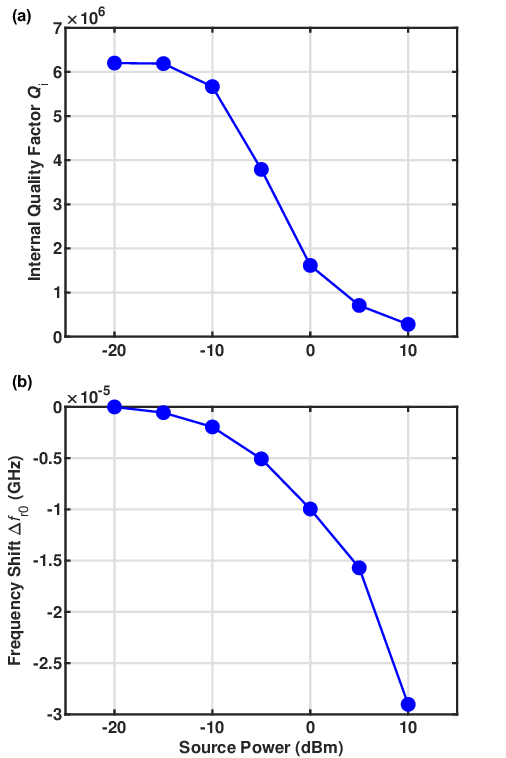}
\caption{\label{fig:phiRM} The fitted (a) internal quality factors, and (b) resonance frequency shifts relative to the maximum fitted frequency at high powers using the $\phi$ rotation method. Both the fitted internal quality factors and resonance frequencies decrease with increasing power, qualitatively agreeing with the readout-power heating model.
}
\end{figure}

On the other hand, the readout-power heating model \cite{de2010readout} predicts that the increasing microwave power induces a higher effective electron temperature, generating excess thermal quasiparticles, and thus decreasing the internal quality factors and resonance frequencies \cite{de2014evidence}. These excess thermal quasiparticles could account for dissipative nonlinearity. However, the fit model we implemented directly extracts the resonance frequencies in the linear regime, $f_{r0}$, which is independent of the power, and $f_r$ is no longer a constant. To directly compare the results with the model prediction, we employed the $\phi$ rotation method \cite{gao2008} to fit the data, and the results are presented in \figref{fig:phiRM}. Here, the resonance frequencies decrease as the power increases. Both the drop in the internal quality factor and frequencies qualitatively agree with the readout-power heating model.

Moreover, the resonators are $\lambda/2$, so the open-ended resonator center strips are isolated from the ground plane, whereas the short-ended ones join the ground plane. This could explain why open-ended resonators exhibit stronger nonlinearity, as it prevents the outdiffusion of thermal quasiparticles into the ground bath. The thermal isolation induced increase in nonlinearity also qualitatively agrees with the readout-power heating model.

\begin{figure}[htb]
    \includegraphics[width=\linewidth]{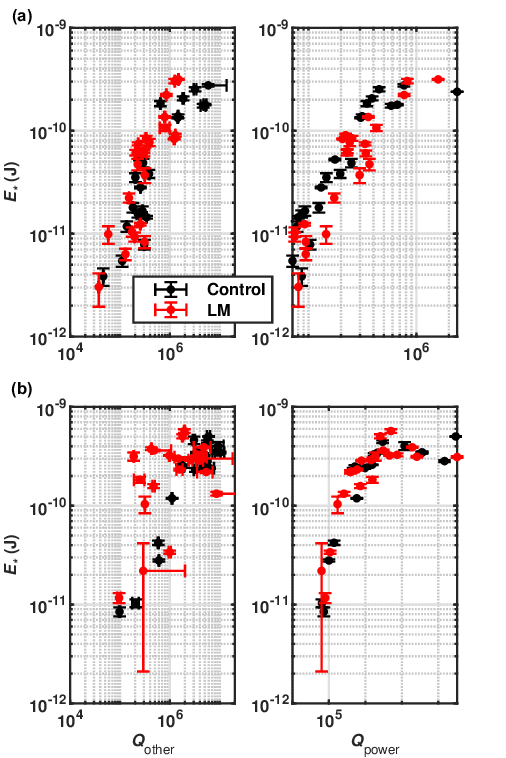}
    \caption{The scaling energy vs. the power-independent and power-dependent quality factor (a) open- and (b) short-ended resonators at 10~dBm source power. The open-ended resonators exhibit stronger monotonic correlation between the scaling energy and both quality factors compared to short-ended ones. The values of $E_*$ are the weighted averages of each resonator, and error bars are calculated from each $95\%$ confidence interval shown in \figref{fig:Eas}(b). The quality factors and $95\%$ confidence intervals are extracted from the fits.}
    \label{fig:Eascorr}
\end{figure}

Besides, we found a nearly monotonic correlation between $E_*$, $Q_\mathrm{other}$, and $Q_\mathrm{power}$ for the open-ended resonators, as shown in \figref{fig:Eascorr}(a). Although $E_*$ and $Q_\mathrm{other}$ should not be dependent, this may be explained by the overlap between the enhanced heating induced loss and onset of TLS saturation manifested by the intersection of the TLS saturation uphill and the power fit downhill shown in \figref{fig:losshi}(b). The caveat is that the fitted $Q_\mathrm{other}$ could be an underestimate if the $\delta_\mathrm{power}$ is more dominant than $\delta_\mathrm{other}$ in the relatively linear regimes and before TLS saturation. This could partially explain the significant percentage of power-independent loss observed at low power, as shown in \figref{fig:loss}(a), but a more thorough investigation is necessary. For short-ended resonators, the correlation is less pronounced, potentially meaning the outdiffusion of the heating quasiparticles induces less heat in the relatively linear regime. It is challenging to disentangle the power-dependent loss at high power from the power-independent loss without incorporating a closed-form physics model into the fitting function. Therefore, we may overestimate the power-dependent loss shown in Figures~\ref{fig:loss}, \ref{fig:aging}, and \ref{fig:losshi} at least for the open-ended resonators. This suggests that an accurate physics model, rather than a phenomenological fit, is required to precisely extract the power-dependent loss.

Overall, the data set quantitatively agrees with the readout-power heating model, and future intermodulation measurements could reveal the detailed nonlinear mechanism \cite{dahm1997theory}.

\section{\label{sec:disc}Discussion}

Here, we have demonstrated the gallium-based liquid metal interconnects bridging coplanar waveguide resonators across discrete modules. Despite the added complexity of inter-module interfaces, the liquid metal interconnects maintained comparable microwave performance to the control resonators at 15~milikelvin, showing negligible degradation even after repeated thermal cycling. Meanwhile, the virtually zero resistance before and after module replacement highlights the potential of liquid metal as a reusable connectivity solution for modular quantum systems. Further, we investigated the underlying physics behind the observed performance deviations at different power levels. Our results suggest that the presence of $\beta$-tantalum, which is supported by XRD and depth-profiling XPS, contributes a significant kinetic inductance fraction, leading to substantial shifts in resonance frequency. Our XPS study further reveals TaN and other interfacial compounds, motivating future studies to pinpoint which materials (or combination) account for the large kinetic-inductance fraction. While our data qualitatively support the readout-power heating model, a more accurate physics model is required to precisely extract the power-dependent loss in high-power regimes. This could also help identify the dissipative contribution of the kinetic inductance nonlinearity.

Together, these results establish liquid metal interconnects as a compelling approach for building modular quantum hardware. They demonstrate not only feasibility but also performance on par with conventional interconnect solutions, paving the way for more scalable and reconfigurable superconducting quantum processors. Despite the promises, several future directions arise.

From an implementation perspective, the immediate next step would be the demonstration of a two-qubit system coupled via liquid metal interconnects. Such a milestone would mark a significant advancement toward fully modular quantum processors. However, the current fabrication process, which relies on hydrochloric acid for oxide and residue removal, introduces manual variability and raises compatibility concerns with aluminum-based Josephson junctions. Therefore, significant engineering effort is required to develop an automated and precise liquid metal printing technique that eliminates the need for chemical cleaning. This would not only improve process reliability but also enhance scalability and integration. Recent efforts have been made to print liquid metal superconducting lumped-element resonators with linewidths close to 10 $\mathrm{\mu m}$ \cite{kreiner2025liquid}, while other proposed printing techniques \cite{ma2023shaping} may be reengineered to fit quantum applications.

With an automated and repeatable liquid metal deposition process in place, future studies on loss mechanisms are likely to yield more conclusive and generalizable insights. A deeper understanding of the electromagnetic behavior of liquid metal interconnects, achieved through a combination of materials characterization, structural imaging, and comprehensive simulation both before and after cryogenic microwave measurements, would provide valuable insights. Such data would benefit both the fundamental study of condensed matter systems involving liquid metal and the continued optimization of liquid metal interconnect designs for quantum applications.

In summary, this work lays the foundation for a new class of modular and reconfigurable superconducting quantum hardware enabled by liquid metal interconnects, with promising directions for both engineering scalability and fundamental discovery.



\begin{acknowledgments}
This work was supported by an IBM-sponsored research Award No.~W2178130. The authors acknowledge the use of facilities at the Boston University Photonics Center and the Harvard University Center for Nanoscale Systems (CNS). CNS is a member of the National Nanotechnology Coordinated Infrastructure Network (NNCI), which is supported by the National Science Foundation under NSF Award No.~ECCS-2025158. The Bruker D2 Phasor XRD was provided by Boston University's Division of Materials Science and Engineering. For XPS measurements, we acknowledge National Science Foundation Grant No.~2216008.
\end{acknowledgments}

\appendix


\section{\label{sec:fab}Fabrication of the liquid metal interconnected module assmbly}

The first step is to pattern the microwave circuit base layer, pre-etch away the metal layer to define the matching shape for the subsequent deep reactive ion etching (DRIE) of the matching edges, all on the top chips. We deposited 200 nm of Ta by physical vapor deposition and then 40 nm of TiN by chemical vapor deposition on roughly 740 $\rm{\mu m}$ high-resisivity silicon in a cluster deposition tool without breaking the vacuum. The silicon wafers were precleaned to remove the native oxides before being sent to the vacuum chamber. All the metal-coated wafers were fabricated and then stored in stock for later use, so they were patterned months after fabrication. We patterned the designed layout of superconducting microwave circuits and the matching edge trench on positive photoresist using a maskless aligner, followed by dry etching using a mixture of $\rm{CHF_3}$ and $\rm{SF_6}$ in an inductively coupled plasma reactive ion etcher. We then stripped the photoresist in a hot Remover 1165 bath. The base layer pattern was now ready.

Next, we patterned the gold pads and gold coating on the back side of the top chips and the matching areas on the carrier chips. After diluting and drying the patterned wafer, we used a bilayer photoresist consisting of a LOR seed layer and a positive tone photoresist to cover the patterned surface. We used a maskless aligner to align and define the gold pads on the two sides of the matching edge trench. In the meantime, we opened up some rectangular areas on the same bilayer photoresist on bare high-resistivity silicon wafers. These areas are designed to be the gold area on the carrier chips, with the same shape as the combined shape of chips A and B after DRIE. We then used an electron-beam evaporator to deposit 20 nm of Ti, followed by 100 nm of gold, without breaking the vacuum. Immediately after the deposition, we coated another layer of photoresist on top of the gold coating on the top chip wafer as a protection layer and then flipped it to coat the backside with 20 nm Ti and 100 nm of gold. This coating serves as the adhesion layer for the backside gallium. After we lifted off the gold-coated photoresist in a Remover 1165 hot bath, we had all the gold patterns.

The last step of fabrication is to etch through the top chip wafers to define the edges of chips A and B. This method enables lithographically defined precision and flexible shapes for alignment, whereas conventional cutting methods, such as dicing, can only cut straight lines. We used a 13-$\rm{\mu m}$ positive photoresist patterned and aligned by the maskless aligner to mask the following DRIE process. After the DRIE, we removed the lingering gold by a quick sonication in Remover 1165 and then stripped the photoresist in the hot bath. We contained the chiplets in a stainless steel mesh strainer and submerged them twice in a bath of acetone, methanol, isopropanol, and deionized (DI) water in succession. We then finished the process with an isopropanol spray and allowed it to dry naturally in the chemical hood. Although visible residues were not always observed under the microscope, the stripping and dilution steps were properly carried out, and any remaining water-soluble residue would be eliminated during the subsequent rinse in the assembly process.

Further, we need to align and assemble chips A and B on the carrier chip and hold them together with gallium. We first selected some chips A and B and inspected the type of resonators under the microscope so that when we combined them, we would have half-wavelength resonators with both ends either open or shorted, not quarter-wavelength resonators. Then, we cleaned the residue, if any, from the previous natural drying process with DI water and HCl. After diluting and drying the chiplets, we dipped the gold surface on the backside into molten gallium on a mildly warm hotplate and then thinned it by gently sliding it on the glassware. We placed the cleaned carrier chips in a diluted HCl solution and then positioned chips A and B with molten gallium on the gold area of the carrier chips. We note that gallium exhibits a significant supercooling effect; therefore, as long as the HCl bath and room temperature are not too cold, it should remain in a liquid state. We used the tweezers to gently pinch chips A and B so that they match the complementary edge shapes. Then, we pressed them to squeeze out any excess gallium, maintaining a relatively flat surface. Then, we diluted the HCl solution with DI water and dried it with nitrogen. We inspected the alignment under the microscope and achieved an alignment accuracy within $\pm$2~$\mathrm{\mu m}$ after several trials and errors. Once we were satisfied with the alignment, we froze the assembled chips in liquid nitrogen, and they will remain solid as long as the processing temperature does not exceed the melting point of gallium.

Finally, we deposited the LM to interconnect the chips A and B. We first used a glass tip to dip into an LM droplet and painted a reservoir of liquid metal hundreds of microns in diameter on a chip with a similar height to the assembled chip. This chip, with a touch of LM, serves as an inkwell for LM. We did not directly take LM from a bulk as the excess LM on the tip could tarnish the chip surface. We used a polyethylene tip with a diameter of approximately 50 $\mathrm{\mu m}$ as a print head, as it would not scratch the surface like glass. We inked the write head in the small LM reservoir and tapped it onto the gold pads. We then used HCl solution and vapor to clean the excess LM residue and reform the LM adhered to the gold pads. As a finite gap is always present between chip A and B, water tends to accumulate in the gap, along with the soluble residue, making the LM interconnects dirty if the DI water flush is insufficient. Thus, a pressured water stream flushed along the joint edges, and then a prolonged nitrogen blow-dry was applied in the same direction to reduce the likelihood of trapping residue. Those LM droplets that adhered well to the gold pads would not be flushed away despite the strong push.

\section{\label{sec:fridge}Microwave measurement apparatus}

\begin{figure}[htb!]
\includegraphics[width=\linewidth]{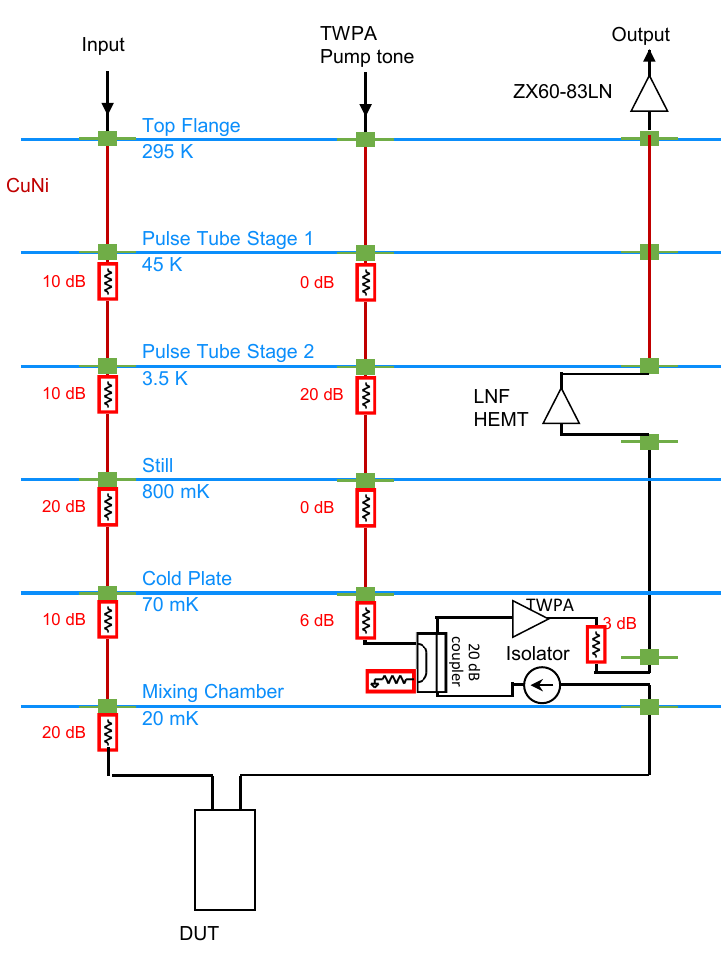}
\caption{\label{fig:fridge} Wiring diagram for the dilution refrigerator. The temperatures of different cooling stages are approximate.
}
\end{figure}

All devices were measured in a Bluefors dilution refrigerator with a base mixing chamber temperature at approximately 15 mK. Twelve independent CuNi input lines and corresponding output lines were used for the input and output signal paths for each chip assembly. A wiring diagram depicting the input and output lines is shown in \figref{fig:fridge}.

A total of $-70$ dB of fixed attenuation was from attenuators at various temperature stages, while a $-10$ dB of temperature-dependent coaxial cable attenuation was measured at room temperature, which roughly translates to $-5$ dB at 4K. Therefore, a total of $-75$ dB attenuation was used when calculating the input power at the transmission lines of the chiplets.

The output signals from the transmission lines were passed through an isolator and then mixed with a traveling wave parametric amplifier (TWPA) pump tone in a coupler at the cold plate stage when the source power is below $-30$ dBm. The TWPA tone was pumped through a separate line with approximately $-26$ dB fixed attenuation. A $-3$ dB attenuator at the cold plate stage and a Low Noise Factory high-electron mobility transistor (HEMT) amplifier at the pulse tube stage 2 were used at the output lines.

The measurements were conducted with a vector network analyzer. Typically, the VNA measurements of resonators spanned between 3 and 8 GHz (which is the approximate maximum range afforded by the TWPA bandwidth), using up to 100,000 points per scan with averaging dependent on the power levels used during the measurement. Power applied to the top of the fridge ranged from +10 to -90 dBm. 

\section{\label{sec:fitmod}Fit model for individual resonance response}

We have observed that the conventional diameter correction method (DCM) fails at high powers, potentially due to nonlinear kinetic inductance. Thus, we implement a correction to the resonance frequency as a function of the circulating current (or stored energy) in the resonator in addition to the DCM.

To begin with, we first briefly review the DCM. As we adopted the ``hanger'' style of resonators coupled to a common TL, the transmission coefficient, $S_{21}(f)$, can be formulated as follows \cite{gao2008,chen2025efficient}:
\begin{eqnarray}
S_{21}(f)=e^{-i 2 \pi f \tau} z(f),
\label{eqn:S21}
\end{eqnarray}
where
\begin{eqnarray}
z(f)= a e^{i \alpha}\left[1-\frac{Q_l / Q_c e^{i \phi}}{1+2 i Q_l\frac{f-f_{r0}}{f_{r0}}}\right]
\label{eqn:zf}
\end{eqnarray}
is the $S_{21}$ at the probe frequency, $f$, after cable delay removal, $f_{r0}$ is the resonance frequency of the resonator in the linear regime, $\tau$ is the cable delay, $a$ and $\alpha$ are the complex modulus and phase of the off-resonance point, respectively, $\phi$ is the complex phase of on-chip impedance mismatch, and $Q_l$ and $Q_c$ are the loaded and coupling quality factors, respectively. After extracting all the fitting parameters, the internal quality factor, $Q_i$, can be calculated by \cite{chen2025efficient,khalil2012}
\begin{eqnarray}
\frac{1}{Q_i} = \frac{1}{Q_l} - \frac{1}{Q_c^{'}}, Q_c^{'} = \frac{\cos{\phi}}{Q_c},
\label{eqn:DCM}
\end{eqnarray}
where $Q_c^{'}$ is the diameter-corrected coupling quality factor. A conventional DCM ends here, and one may use the fitting parameters to conduct further analysis. To incorporate current-dependent nonlinearity in kinetic inductance, we can write the kinetic inductance of the resonator, $L_k(I)$, as a function of resonator circulating current, $I$ \cite{swenson2013operation,zmuidzinas2012superconducting}:
\begin{eqnarray}
L_k(I)=L_{k0}\left[1+\left(\frac{I}{I_*} \right)^{2}+...\right]\,
\label{eqn:KII}
\end{eqnarray}
where $L_{k0}$ is the kinetic inductance of the resonator at zero circulating current limit, $I_*$ is the scaling current of the quadratic nonlinearity \cite{swenson2013operation}, and is expected to be on the order of the critical current of the superconducting material of the resonator. Now, the resonance frequency, $f_r$, is a function of $I$, and the nonlinear fractional resonance frequency shift is given by

\begin{equation}
\begin{split}
\delta x = \frac{f_r-f_{r0}}{f_{r0}} = -\frac{1}{2}\frac{L_k(I)-L_{k0}}{L} \\=-\frac{\alpha_LI^2}{2I_*^2}=-\frac{E}{E_*}
\end{split}
\label{eqn:detx}
\end{equation}

where the $\alpha_L$ is the kinetic inductance, $E$ is the stored energy in the resonator, and the scaling energy $E_* \propto L_kI_*^2/\alpha_L$ is expected to be on the order of the condensation energy of the superconductor if $\alpha_L\approx1$ \cite{swenson2013operation}. Then, the $f_r$ can be written as

\begin{eqnarray}
f_r=f_{r0}\left(1-\frac{\alpha_LI^2}{2I_*^2}\right)=f_{r0}\left(1-\frac{E}{E_*}\right).
\label{eqn:fr}
\end{eqnarray}

and the modulus square of the difference between complex $S_{21}$ at the probe frequency and off-resonance point after cable delay removal, $\lvert z(f)-z_{\rm{off}}\rvert^2$ is proportional to $I^2$ (or $E$), according to the geometric relationship in complex plane \cite{dai2022new}.  \eqnref{eqn:fr} can be rewritten as

\begin{equation}
\begin{split}
f_r=f_{r0}(1 - \beta\lvert z(f)-z_{\rm{off}} \rvert^2) \\ = f_{r0}(1 - \beta\lvert z(f)-a e^{i \alpha}\rvert^2),
\label{eqn:frb}
\end{split}  
\end{equation}

where the $\beta$ is a dimensionless fitting parameter.
Substituing \eqnref{eqn:frb} into \eqnref{eqn:zf}, we have our nonlinear model:

\begin{multline}
z(f)= \\ ae^{i \alpha}\left[1-\frac{Q_l / Q_c e^{i \phi}}{1+2 i Q_l\left(\frac{f}{f_{r0}(1 - \beta\lvert z(f)-a e^{i \alpha}\rvert^2)}-1\right)}\right].
\label{eqn:zfn}
\end{multline}

However, a direct nonlinear least-squares fitting of this model often fails, as no direct closed-form minimization can be found, and iterative methods can easily lead to local minima if poor initial guesses are provided. The step-by-step fitting procedure is explained in \secref{sec:proc}.

\section{\label{sec:proc}Fitting procedure for extracting resonator parameters}

Here, we discuss how to effectively extract the fitting parameters of interest without getting trapped in local minima.

\subsection{\label{sec:delay} Cable delay removal}

\begin{figure}[htb!]
  \includegraphics[width=\linewidth]{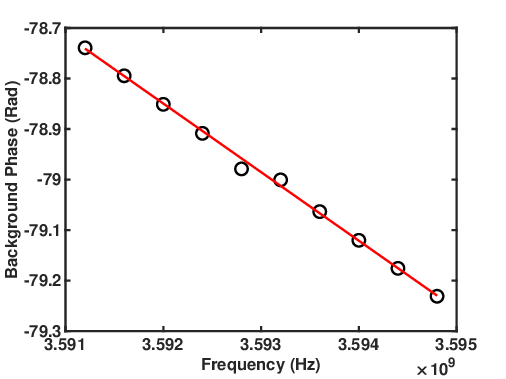}
  \caption{A linear fit in the background $S_{21}$ phase response near the resonance frequency.}
  \label{fig:widerange}
\end{figure}

The first step is to remove the cable delay term in \eqnref{eqn:S21} so that the $S_{21}$ of the resonator restores a circular shape in the complex plane. One could linearly fit the phase background as shown in \figref{fig:widerange}, and extract the slope as $-2\pi\tau$.

\begin{figure*}
  \includegraphics[width=\linewidth]{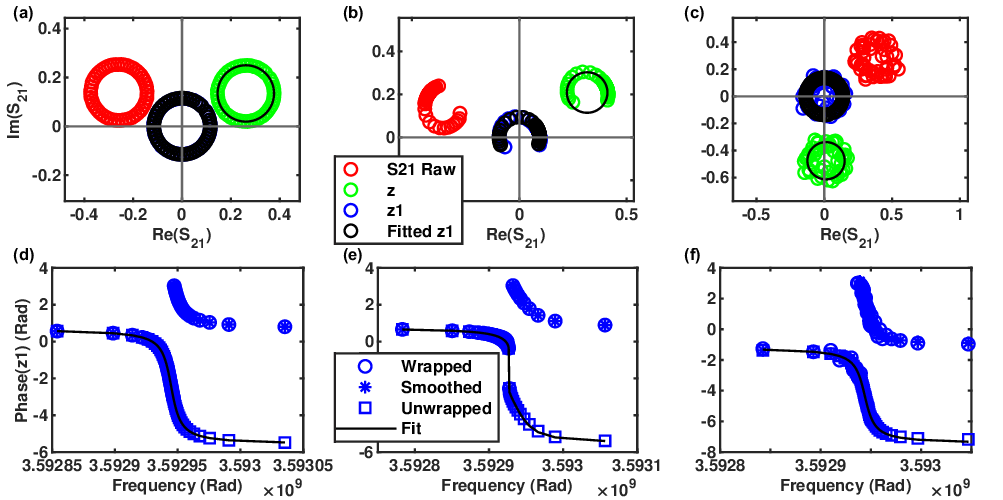}
  \caption{\label{fig:complexfine} Fitting procedure for (a) linear, (b) nonlinear, and (c) noisy resonators, and specifically their corresponding phase fit (d), (e), and (f), respectively, before the direct fit to \eqnref{eqn:zfn}. (a)-(c) The complex $S_{21}$ data at different stages of the fitting procedure. The raw data is plotted in red circles. After the cable delay term, $e^{-i 2 \pi f \tau}$, is removed, the data is reduced to $z(f)$ plotted in green circles. A circle fit in $z(f)$ is performed and plotted in the black solid curve. Then, $z(f)$ is subtracted by the circle center, which equals $z_1(f)$ and is plotted in blue circles. A circularization of $z_1(f)$ is performed by $z_{\rm{1,circ}}(f)=R\cdot e^{i\angle z_1(f)}$, where $R$ is the extracted radius from the previous circle fit. The phase fit results are plotted in black circles. (d)-(f) The phase fit using $\angle z_1(f) = 2\operatorname{arctan}(2Q_l(1-f/f_r))-\theta$, where $f_r = f_{r0}(1 - \beta \lvert z_1(f)-z_{\mathrm{1,off}}\rvert^2)$. The wrapped phase data is first plotted in blue circles. Then, depending on the type of resonator, different types of post-processing are implemented before the phase fit. For very high power data, for example, the data in (e), an algorithm with an empirical threshold is implemented to detect phase wrapping jumps, as the sudden drop in the phase angle due to nonlinear bifurcation can be falsely identified as a wrapping jump by a conventional unwrap function. For very low power data like the data in (f), a Gaussian filter is implemented on $\sin{(\angle z_1(f))}$ and $\cos{(\angle z_1(f))}$, and they are combined as the smoothed phase (blue asterisks), $\rm{atan2}(\sin,\cos)$, before the implementation of the conventional unwrap function. Finally, the phase fit is performed using the post-processed data (blue squares), and the extracted parameters are used as initial guesses for the subsequent fitting.
}
\end{figure*}

Although it is possible to treat the extracted $\tau$ as an initial guess of the circle fit \cite{probst2015efficient} or the direct nonlinear fit to \eqnref{eqn:S21}, we find that a slight change in $\tau$ could lead to significant deviation from the global fitting optimum. For the robustness of the fitting algorithm, we chose to fix $\tau$ for the subsequent fitting procedure. After the cable delay removal, the raw $S_{21}$ data (red) is reduced to $z(f)$ (green), as shown in \figref{fig:complexfine}(a)(b)(c).

\subsection{\label{sec:circle}Circle fit}

We adopted the algebraic least squares fitting of circles by Chernov and Lesort \cite{chen2025efficient}, which is also discussed by \cite{gao2008,probst2015efficient}. After $z(f)$ is calculated from the last step, the parametrization of a circle is as follows \cite{gao2008,probst2015efficient,chernov2005least}:
\begin{equation}
\begin{split}
A(x^2+y^2)+Bx+Cy+D=0 \\ x= \operatorname{Re}(z), y=\operatorname{Im}(z),
\label{eqn:paracirc}
\end{split}
\end{equation}
which is subject to the constraint, $B^2+C^2-4AD=1$. The objective function to be minimized is given by
\begin{eqnarray}
\mathcal{F}(A, B, C, D)=\sum_{i=1}^n\left(A z_i+B x_i+C y_i+D\right)^2.
\label{eqn:objfun}
\end{eqnarray}
In the matrix form, $\mathcal{F}=\mathbf{A}^T \mathbf{M} \mathbf{A}$ with $\mathbf{A}=(A,B,C,D)^T$, and is subject to the constraint $\mathbf{A}^T\mathbf{B}\mathbf{A}=1$, where
\begin{eqnarray}
\mathbf{M}=\left(\begin{array}{cccc}
M_{z z} & M_{x z} & M_{y z} & M_z \\
M_{x z} & M_{x x} & M_{x y} & M_x \\
M_{y z} & M_{x y} & M_{y y} & M_y \\
M_z & M_x & M_y & n
\end{array}\right) ,
\label{eqn:pmatrix}
\end{eqnarray}
and
\begin{eqnarray}
\mathbf{B} = \begin{pmatrix}
0 & 0 & 0 & -2 \\
0 & 1 & 0 & 0 \\
0 & 0 & 1 & 0 \\
-2 & 0 & 0 & 0
\end{pmatrix}.
\label{eqn:Bmatrix}
\end{eqnarray}
Here, $M_{ij}$ is the moment of data, e.g., $M_{xx}=\sum_{i=1}^nx_i^2$, $M_{xy}=\sum_{i=1}^nx_iy_i$. Then, a Lagrange multiplier, $\eta$, is introduced to minimize $\mathcal{F}$ subject to the mentioned constraint so that
\begin{eqnarray}
\mathcal{F}_*=\mathbf{A}^T \mathbf{M} \mathbf{A} - \eta (\mathbf{A}^T\mathbf{B}\mathbf{A} - 1).
\label{eqn:Mmin}
\end{eqnarray}
Differentiating \eqnref{eqn:Mmin} with respect to $\mathbf{A}$ gives
\begin{eqnarray}
\mathbf{M}\mathbf{A}-\eta \mathbf{B} \mathbf{A} = 0,
\label{eqn:Mderiv}
\end{eqnarray}
and $\eta$ can be solved by
\begin{eqnarray}
\mathrm{det}(\mathbf{M}-\eta \mathbf{B})=0.
\label{eqn:Mdet}
\end{eqnarray}
Once $\mathbf{M}$ is calculated and $\eta$ is solved numerically, the vector $\mathbf{A}$ can be obtained. The center coordinates and radius of the circle are given by
\begin{eqnarray}
x_c & = & -\frac{B}{2A},
\\
y_c & = & -\frac{C}{2A},
\\
R & = & \frac{1}{2 \lvert A \rvert} \sqrt{B^2+C^2-4AD}=\frac{1}{2\lvert A \rvert}.
\label{eqn:circpara}
\end{eqnarray}
The results of circle fits are plotted in black solid curves among $z(f)$ data points (green circles) in \figref{fig:complexfine}(a)(b)(c).

We can map $z(f) \rightarrow z_1(f)$ by
\begin{eqnarray}
z_1(f) = z(f) - z_c,
\label{eqn:z1}
\end{eqnarray}
where $z_c = x_c + iy_c$. The centered $z_1(f)$ (blue circles) is shown in \figref{fig:complexfine}(a)(b)(c).

\subsection{\label{sec:pha}Phase fit}

Before the phase fit is performed, a phase unwrapping is required. A conventional unwrap function is directly implemented on data acquired at medium power, since the data has little phase noise and the resonance response is linear. The results of unwrapping are plotted in blue squares in \figref{fig:complexfine}(d). The same function can be used for low-power data (\figref{fig:complexfine}(f)), but sometimes the noise fluctuations in phase can be detected as phase wrapping. Thus, a denoising process is applied before unwrapping. Here, we applied a Gaussian filter on $\sin$ and $\cos$ components separately, assuming both $\sin{(\angle z_1)}$ and $\cos{(\angle z_1)}$ are continuous. This assumption is true for resonance responses before the nonlinear bifurcation at high power, as such a bifurcation can induce a sudden drop in phase, as shown in \figref{fig:complexfine}(e). Since we only apply the filter on low-power data, the assumption holds. The advantage of denoising the $\sin$ and $\cos$ components separately is that we can avoid averaging out the data before and after wrapping jumps. After the filter is applied, we recombined the filtered components as $\angle z_\mathrm{1,filtered} = \operatorname{atan2}(\sin_\mathrm{filtered}(\angle z_1),\cos_\mathrm{filtered}(\angle z_1))$, plotted in blue asterisks in \figref{fig:complexfine}(f), and then we unwrapped the filtered data with conventional unwrap function. For highly nonlinear responses inducing bifurcation, we defined an asymmetric empirical threshold for phase wrap detection. As the frequency sweeps from low to high, a drop in phase angle appears when the bifurcation occurs, and thus, the phase wrap detection should tolerate a large drop while tightening the threshold for an increase in phase.

Next, we can perform the phase fit with the unwrapped phase data. The model we used to fit in the unwrapped $\angle z_1(f)$ is
\begin{multline}
\angle z_1(f) = \\ 2\operatorname{arctan} \left [2Q_l \left(1-\frac{f}{f_{r0}(1 - \beta\lvert z_1(f)-z_{\mathrm{1,off}}\rvert^2)} \right) \right]-\theta.    
\label{eqn:pha}
\end{multline}
Here, we have $Q_l$, $\beta$, $f_{r0}$, $\lvert z_\mathrm{1,off} \rvert$, $\angle z_\mathrm{1,off}$, and $\theta$, six fit parameters, where $z_{\mathrm{1,off}} = \lvert z_\mathrm{1,off} \rvert e^{i\angle z_\mathrm{1,off}}$. We note that $\beta$ and $z_\mathrm{1,off}$ are loosely correlated as different $\beta$ and $z_\mathrm{1,off}$ values could yield similar $\beta\lvert z_1(f)-z_{\mathrm{1,off}}\rvert^2$. Thus, instead of letting both the modulus and angle of $z_\mathrm{1,off}$ as free fit parameters, we fix the modulus as the previously extracted circle radius, R. Further, we leverage the geometric relation between the off-resonance point angle and $\theta$ \cite{chen2025efficient}, so that $\angle z_\mathrm{1,off} = \pi - \theta$. Now, we reduce the number of fit parameters to four, and \eqnref{eqn:pha} reads as follows:

\begin{widetext}
\begin{equation}
\angle z_1(f) = \\
2\operatorname{arctan}\left[2Q_l \left(1-  \frac{f}{f_{r0}(1 - \beta\lvert z_1(f)-R\cdot e^{i(\pi-\theta)}\rvert^2)} \right) \right]  -\theta.
\label{eqn:phap}
\end{equation}
\end{widetext}

We set the $R$ and the corresponding $z_1(f_i)$ as the problem-dependent parameters. One additional note on the input $z_1(f_i)$ is that the data can exhibit circle distortion at very high power, as shown in \figref{fig:widerange}(b), or become noisy at low power, as shown in \figref{fig:complexfine}(c), which could add to the noise in the $\beta\lvert z_1(f)-R\cdot e^{i(\pi-\theta)}\rvert^2$ term and lead to abnormal extraction of $\beta$ and other parameters in the phase fit. To circumvent the issue, we performed a circularization of $z_1(f)$, which is plotted in blue squares in \ref{fig:complexfine} (a)(b)(c), and is defined by $z_{\mathrm{1,circ}}(f)=R\cdot e^{i\angle z_1(f)}$. Therefore, the nonlinear least-squares fit becomes:
\begin{eqnarray}
\operatorname{min}\left \{ \sum_{i=1}^n \left[\angle z_1(f_i) -  \angle z_\mathrm{1,fit}(f_i,z_\mathrm{1,circ}(f_i),R)\right]^2 \right \}.
\label{eqn:minpha}
\end{eqnarray}

For any nonlinear least-squares fit, the appropriate choices of initial guesses of the fit parameters are essential to circumvent local minima and divergence. Here, we briefly discuss how we choose the appropriate initial values. The initial value of the $Q_l$ can be defined as $f_\mathrm{peak} / \Delta f$, where $\Delta f$ is the full width of the half maximum of the resonance peak, and $f_\mathrm{peak}$ is the frequency at the peak, which can be a proper initial value for $f_{r0}$. The average phase angle of the $z_1(f)$ data at the lowest and highest frequencies can be the initial guess for the off-resonance phase angle, $\pi-\theta$, assuming the measurement covers at least a few linewidths of the resonance response. Lastly, $\beta$ is a measure of nonlinearity. It is correlated with the resonance circle size, so we can set a dynamic initial guess as $\beta_0P_g/R$, where $\beta_0$ is an empirical constant depending on the studied system, and $P_g$ is the input power at the transmission line.

The fit results are plotted in black solid curves in \figref{fig:complexfine}(d)(e)(f). We are aware of the artifacts introduced by all the data post-processing for a more robust fit, so we only use the results of the phase fit as the initial values of the subsequent fitting procedure.

\subsection{\label{sec:zfit}Direct complex $z(f)$ fit}

\begin{figure*}[htb!]
\includegraphics[width=\linewidth]{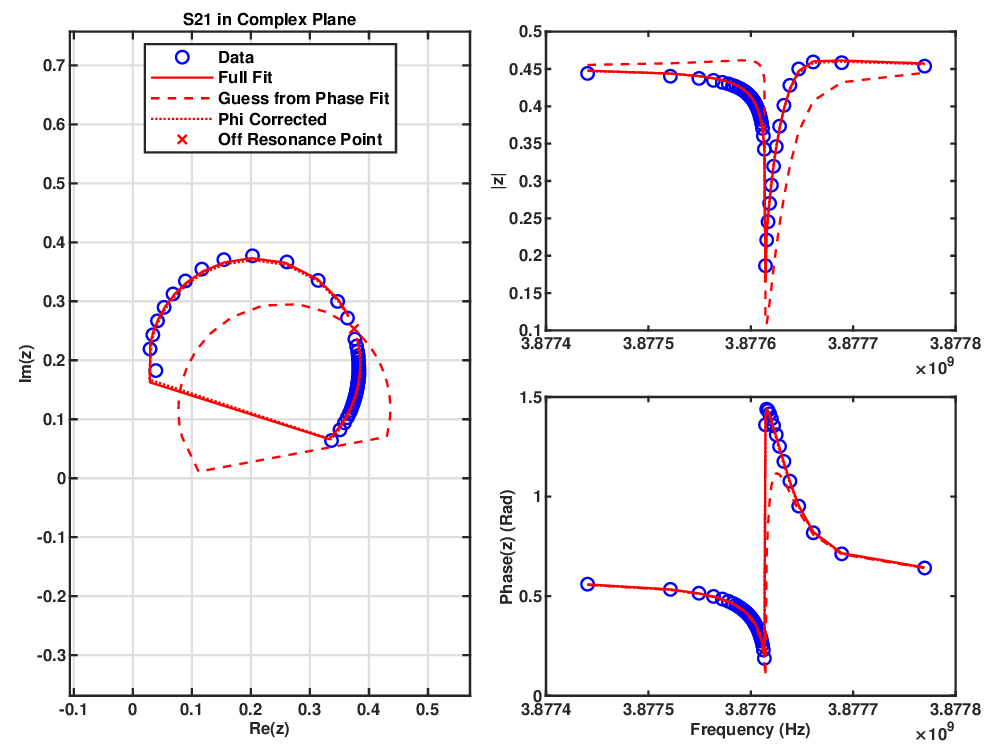}
\caption{\label{fig:zfit} The step-by-step fit results from phase fit (dashed line), $\phi$ correction (dotted line), and the final direct fit (solid line) to a highly nonlinear resonance response after cable delay removal in the complex plane (left), modulus (top-right), and phase (bottom-right) of $z(f)$. We observed that although the guesses of other parameters from the phase fit are close enough to the final estimates, depicted by the almost overlapping phi correction (dotted line) and final estimates (solid line), the $\phi$ guess from $\phi = \alpha + \theta - \pi$ are not very reliable, depicted by the deviation of the phase fit guess (dashed line) from the data. We could attribute the difference to the impedance mismatch at the quarter-wavelength of the resonator due to the change in the geometry, but a more rigorous investigation is required to conclude the root cause.
}
\end{figure*}

Finally, we can use the fit results of the previous circle fit and phase fit to derive the initial values of fit parameters of $z(f)$, defined in \eqnref{eqn:zfn}. The estimated off-resonance point $ae^{i\alpha}=R\cdot e^{i(\pi-\theta)}+z_c$, and further, the estimated $Q_c=Q_la/(2R)$. Usually, these two guesses are very close to the best estimates of the fit parameters. However, we want to emphasize that, although a geometric constraint $\phi = \alpha + \phi - \pi$ should be valid \cite{chen2025efficient}, we often find that the best estimate of $\phi$ has more or less deviation from the initial guess derived from such a relation (\figref{fig:zfit}), inducing a poor convergence. We can attribute the difference to the impedance mismatch at the quarter-wavelength of the resonator due to the change in geometry, but a more rigorous investigation is required to determine the root cause. To correct this, we first estimated $\phi = \alpha + \phi - \pi$ and then fixed all the other fit parameters in \eqnref{eqn:zfn} to perform a nonlinear least-squares fit for a $\phi$ correction. We used the $\phi$ from the previous fit to derive the estimated $Q_i=1/(1/Q_l-\cos{\phi}/Q_c)$, since it is more straightforward to calculate the confidence intervals of $Q_i$ from the fit residuals than from the propagation of errors of the correlated parameters. Additionally, to avoid involving complex number residual without omitting either real or imaginary part of the $z(f)$, we construct the nonlinear least-squares fit as

\begin{equation}
\begin{split}
\operatorname{min}\left \{ \sum_{i=1}^n \left[0 - \lvert z(f_i) -  z_\mathrm{fit}(f_i,z(f_i))\rvert\right]^2 \right \} \\ = \operatorname{min}\left \{ \sum_{i=1}^n \lvert z(f_i) -  z_\mathrm{fit}(f_i,z(f_i))\rvert^2 \right \},
\label{eqn:minzfn}
\end{split}
\end{equation}

using all the previously estimated initial values. Here, we take the modulus of the difference between the measured and fitted $z(f)$ as a new function, where both the real and imaginary parts are still minimized, and all the dependent variable values of the new function become zero. Now, the residuals become real numbers.

We avoid the use of robust nonlinear regression methods like the bisquare function, as we observed that the noise is mainly random, and outliers merely appear. Moreover, a robust algorithm sometimes weights the off-resonance points more than the resonance peak, especially when an incomplete resonance circle appears in a highly nonlinear regime. The $\phi$ correction and the fit results are usually close enough, as shown in \figref{fig:zfit}, where the dotted lines are almost completely covered by the solid lines. All the mentioned steps constitute the fitting procedure, and the final estimates are used for further analyses.

\section{\label{sec:nlin} Extraction of the nonlinear parameters}

As mentioned earlier, we distinguish between linear and nonlinear regimes using the nonlinearity parameter. Although $\beta$ in \eqnref{eqn:frb} is a measure of the nonlinearity, it is convenient to compare material-dependent quantities like $E_*$ and $I_*$ \cite{shu2021nonlinearity}, and to quantify the extent of the bifurcation by the nonlinearity parameter \cite{swenson2013operation}:

\begin{eqnarray}
a_n = \frac{2Q_l^3}{Q_c}\frac{P_g}{\omega_rE_*},
\label{eqn:ana}
\end{eqnarray}

where $\omega_r$ is the angular resonance frequency as a function a $I$. The $P_g$ can be estimated by the input power of the network analyzer and circuit attenuation, and all the other parameters except $E_*$ are estimated by the previous fit. Therefore, we need to determine $E_*$ to derive $a_n$.

\begin{figure}
  \includegraphics[width=\linewidth]{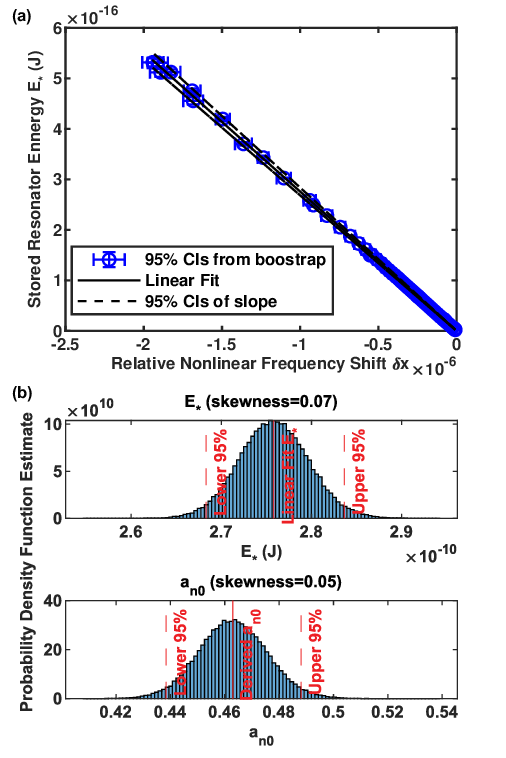}
  \caption{Derived values and $95\%$ confidence intervals of $E_*$ and $a_n$ (a) A linear regression through the origin of the stored energy in the resonator vs. the fractional resonance frequency shift. The $95\%$ confidence intervals of both derived $x$ and $y$ values and the slope are estimated by the 2.5th and 87.5th percentiles of the bootstrap distribution of $10^5$ times concurrent ramdon resampling of all the fit parameters in \eqnref{eqn:minzfn}, assuming that they are subject to $t$ distributions with the standard errors and degrees of freedom derived from the fit. (b) The bootstrap distribution of $E_*$ and $a_n$ derived from the resampled fit parameters. The left and right dashed lines denote the 2.5th and 97.5th percentiles of the distribution, respectively, and the solid lines represent the values derived from the best estimates of the fit parameters. The y-axis is the probability density function estimated from the histogram.}
  \label{fig:E_delx_an}
\end{figure}

We can simply rewrite \eqnref{eqn:detx} as follows:
\begin{eqnarray}
E=-E_*\delta x.
\label{eqn:Eas}
\end{eqnarray}
One means to determine $E_*$ is to derive an array of $E$ and $\delta x$ and then extract the slope of linear regression in \eqnref{eqn:Eas} as $-E_*$. 
From \eqnref{eqn:frb}, we can calculate an array of $fr$, then substitute them into \eqnref{eqn:detx} to calculate $\delta x$. For $E$ array, we need to first define the fractional detuning of the probe frequency as
\begin{eqnarray}
x = \frac{f-f_r}{f_r}.
\label{eqn:x}
\end{eqnarray}
According to the conservation of power \cite{swenson2013operation}, the stored energy in the resonator is found to be
\begin{eqnarray}
E = \frac{2Q_l^2}{Q_c}\frac{1}{1+4Q_l^2x^2}\frac{P_g}{\omega_r}.
\label{eqn:E}
\end{eqnarray}
After the array of $E$ is calculated, we can solve the matrix equation $\mathbf{\delta x}(-E_*)=\mathbf{E}$, where $\mathbf{\delta x} = (\delta x_1,\delta x_2,...,\delta x_n)^T$, $\mathbf{E}=(E_1,E_2,...,E_n)^T$, and $n$ is the number of data points. This is equivalent to linear regression through the origin, and the result is shown in \figref{fig:E_delx_an}(a). Then, we substitute the point estimate of $E_*$ into \ref{eqn:ana} to calcuate $a_{n0}=a_n(f_0)$. The point estimates of $E_*$ and $a_{n0}$ within a bootstrap distribution are shown in \figref{fig:E_delx_an}(b).

 In the figure, we also show the $95\%$ confidencne intervals of the individual point estimate of $E$ and $\delta x$, along with the intervals of the slope. The lower and upper bounds of these intervals are the 2.5th and 97.5th percentiles of a bootstrap distribution calculated from $10^5$ times concurrent random resampling of all the fit parameters in \eqnref{eqn:minzfn}. Here, we discuss the bootstrapping process.
 
 We note that $E_*$ and $a_{n0}$ are all derived values, and the parameters used to calculate them are point estimates from the previous nonlinear least-squares fitting, meaning all of them possess some uncertainties. As we have mentioned in \secref{sec:hipwr}, our derived $E_*$ values could be several orders of magnitude larger than the condensation energy estimate of the center strip of the resonators. Thus, we need to scrutinize the process to ensure that the confidence intervals of $E_*$ and $a_{n0}$ are close enough to the point estimates, even considering the error propagation. We ended up adopting the bootstrapping to estimate the confidence intervals discussed below.

 The algorithm to fit \eqnref{eqn:minzfn} includes the confidence intervals of all the fit parameters, derived from regression residuals, the Jacobian matrix, and degrees of freedom. Thus, we can reconstruct a $t$ distribution based on these values and resample random numbers subject to the distribution: $p_\mathrm{rnd}=p_0+\mathrm{SE}\cdot t_\mathrm{rnd}(\mathrm{DoF})$, where $p_0$ and $\mathrm{SE}$ are the point estimate and standard error of the fit parameter, respectively, $t_\mathrm{rnd}(\mathrm{DoF})$ is a random number subject to $t$ distribution with a degree of freedom of the fit. We take all the resampled fit parameters and substitute them into \eqnref{eqn:detx} and \eqnref{eqn:E} to calculate $\delta x$ and $E$, perform a linear regression through the origin to extract $E_*$, then calculate $a_{n0}$, and repeat the process $10^5$ times. Here, we have the bootstrap distribution for every derived value, and we take the 2.5th and the 97.5th percentiles of each distribution as the lower and upper bounds of the $95\%$ confidence intervals, respectively. The caveat here is that we assume the independence of the random sampling for each fit parameter and treat $f$, $z(f)$, and $P_g$ as constants, which is still a simplified picture.
 
 We have also tried to calculate the confidence intervals of every $E$ and $\delta x$ point estimate using error propagation of independent variables, then performed both robust linear regression, omitting the errors in point estimates, and Deming regression, incorporating the errors in both $E$ and $\delta x$ values, but both regressions almost always yield a narrower condidence interval compared to the one derived from the mentioned bootstrapping process. Thus, we chose to report the confidence intervals by bootstrap distribution to avoid significantly underestimating the uncertainty.

\section{\label{sec:XRD}X-ray diffraction of the metalization layer}

We performed X-Ray diffraction on the metalized silicon wafer at room temperature and found characteristic $\beta$-Ta, as shown in \figref{fig:XRD} \cite{myers2013beta,abadias2019elastic}. The materials and crystal plane Miller indices (hkl) were corroborated by the literature, and the lattice spacing $d_\mathrm{hkl}$ was determined using Bragg’s law:
\begin{eqnarray}
n\lambda = 2 d_\mathrm{hkl} \sin(\theta),
\label{eqn:bragg}
\end{eqnarray}
and the (002) and (004) planes were used to extract the c-lattice parameter:
\begin{eqnarray}
c = d_{001} = 2 d_{002} = 4d_{004}.
\label{eqn:cpara}
\end{eqnarray}
The pseudo-Voigt fitting results are summarized in \tabref{tab:pvfit}. The pseudo-Voigt fit is a mixture of a Gaussian and Lorentzian distribution, where $\eta$ is the mixing parameter. The measured $\beta$-Ta (002) and (004) crystallographic planes have spacings of 0.268 and 0.135 nm, respectively, yielding a c-lattice parameter of 0.536-0.54 nm. These values are larger than the single-crystal $\beta$-Ta (002) and (004) Bragg diffractions measured by Arakcheeva et al. (0.265 and 0.133 nm at 293 K) \cite{arakcheeva2003commensurate}. Assuming uniform elastic strain, the out-of-plane strain can be determined by \cite{hidnert1929thermal,knepper2007coefficient}
\begin{eqnarray}
\epsilon_\mathrm{zz} \equiv \frac{c-c_0}{c_0}
= \frac{d_{002}^\mathrm{meas} - d_{002}^\mathrm{bulk}}{d_{002}^\mathrm{bulk}}.
\label{eqn:epszz}
\end{eqnarray}
It was determined that $\beta$-Ta has an $\epsilon_{\mathrm{zz}}$ of +0.98-1.17\% using \eqnref{eqn:epszz}. This strain could result from internal stresses intrinsic to the deposition method.

\begin{table}[htb!]
\caption{\label{tab:pvfit} Proposed compositional and crystallographic planes from the X-ray diffraction data shown in \figref{fig:XRD} were estimated from pseudo-Voigt fitting. The fitted center $2\theta$ value was used to predict the Bragg plane spacing, pseudo-Voigt mixing coefficient ($\eta$), and the coefficient of determination ($R^2$) of the fit.}
\begin{ruledtabular}
\begin{tabular}{cccccc}
Material & ($\mathrm{hkl}$) & $2\theta$ (\textdegree) & $d_\mathrm{hkl}$ (nm) & $\eta$ & $R^2$\\
\hline
$\beta$-Ta & 221 & 30.1273 & 0.2964 & 0.1902 & 0.9583\\
$\beta$-Ta & 002 & 33.4045 & 0.2680 & 0.2823 & 0.9979\\
c-TaN & 220 & 61.9634 & 0.1496 & 0.0984 & 0.9710\\
c-TiN & 220 & 62.2340 & 0.1491 & 0.4675 & 0.8529\\
$\beta$-Ta Cu $\mathrm{K\alpha}_{1}$& 004 & 69.3937 & 0.1353 & 0.9678 & 0.9946\\
$\beta$-Ta Cu $\mathrm{K\alpha}_{2}$& 004 & 69.5885 & 0.1353 & 0.8696 & 0.9971\\
\end{tabular}
\end{ruledtabular}
\end{table}

Additional Bragg diffraction signatures were observed and are hypothesized to be cubic TaN (220) and cubic TiN (220). It is hypothesized that TaN (220) is cubic as it is at the interface between $\beta$-Ta, a tetragonal crystal, and TiN, a face-centered cubic (FCC) crystal \cite{molodovskaya1975neutron,lei2014synthesis} and facilitates the transition to the c-TiN adlayer within the heterostructured crystal. The proportion of c-TaN (220) is expected to be limited to the interface, which explains the low intensity in comparison to the $\beta$-Ta; the angle of c-TaN (220) Bragg diffraction is corroborated by \cite{shen2011effect,kozlowska2009properties}. Overall, more work could be conducted on engineering the stress in the $\beta$-Ta film if one needs to fine-tune the kinetic inductance.

The X-ray measurements were conducted using a Bruker D2 Phasor benchtop XRD in 1D mode in the Bragg angle domain of $25\text{--}85^\circ$ with increments of $0.0202^\circ$ and 0.3 seconds per increment. The X-ray source was operated at 300 W (30 kV, 10 mA) using a 0.2 mm divergence slit.

\section{\label{sec:XPS}X-ray photoelectron spectroscopy sputter depth profiling}

Here, we discuss the results, experiment, and analysis method of XPS sputter depth profiling from the surface TiN encapsulation layer to the top of the Ta base layer.

\subsection{\label{sec:XPSresult} Depth profiling analysis}

\begin{figure*}[!htb]
\centering
\includegraphics[width=\textwidth]{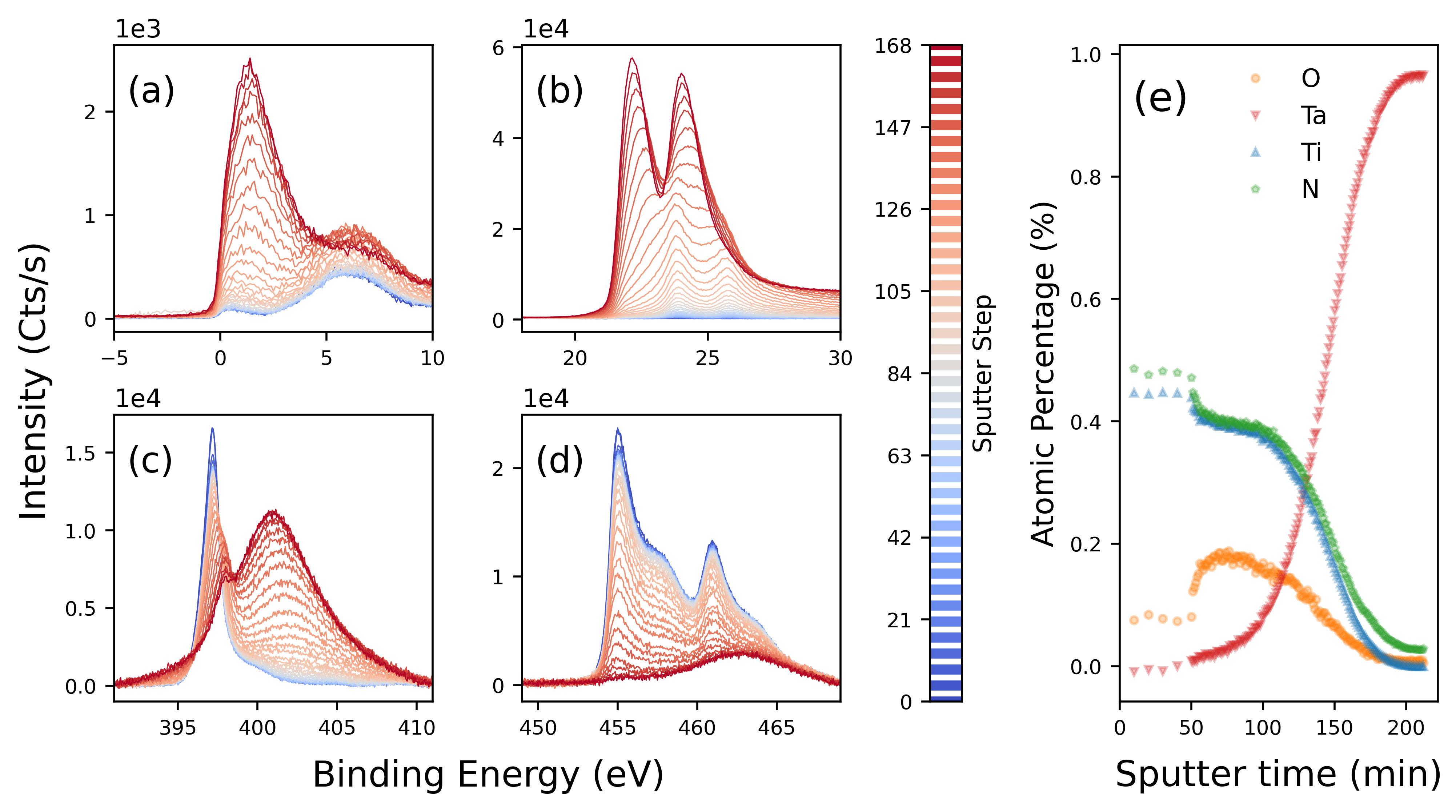}
\caption{(a-d) XPS spectra for (a) Valence band, (b) Ta 4f, (c) N 1s, and (d) Ti 2p core levels. A Tougaard background is subtracted from spectra in (c,d) for clarity. 
Every 10th sputter step is displayed. 
The dark blue line corresponds to the initial sample and proceeds to dark red for the last dataset. 
The first 5 sputter steps are 10 min each, and then it proceeds by 1 min steps. The stoichiometry as determined at each point in the sputtering process is in panel (e).}
\label{fig:XPSN1}
\end{figure*}

The cascades of core level spectra are presented in \figref{fig:XPSN1}.
The spectral datasets are colored, as sputtering proceeds, from the initial sample with deep blue to the sample after 167 min of sputtering in dark red.
The valence band (a) and Ta 4f (b) are presented ``as is'', and the N 1s (c) and Ti 2p (d) have been plotted with the Tougaard background subtracted for visual clarity. 
The increase in the background is due to the disorder induced by sputtering. 
The valence band demonstrates that all spectra are calibrated to the Fermi level for each sputtering step.
\figref{fig:XPSN1}(a) demonstrates the continued presence of the fermi edge, indicating the majority of the sample is metallic during the entire sputtering process: at the fermi level are the partially filled Ti 3d states while hybridized Ti 3d - N 2p states comprise the band at slightly higher energy between 5 to 7 eV \cite{edaS&IAXrayPhotoelectronSpectroscopybased2022}. 
The increase, then decrease of the feature at 6 eV is concurrent with the revealing and eventual sputter-ablation of the TaN and Ta layers consistent with previous studies of N implementation in Ta films \cite{arranz2000tantalum,arranz2005composition}. 

The presence of intermediate tantalum nitride phases can be seen via the change in the appearance and shift of the Ta 4f spectra shown in \figref{fig:XPSN1}(b) and the decomposition in \figref{fig:XPSN2}(a), where the appearance of the $\mathrm{Ta^0}$ component is preceded by the appearance and disappearance of a cubic TaN phase.
This first TaN phase is followed by a $\mathrm{Ta_2N}$ phase, then a $\mathrm{TaN_{0.05}}$ phase, before the $\mathrm{Ta^0}$ phase becomes dominant. The rise of the $\mathrm{Ta^0}$ phase is concomitant with the decrease in intensity of the nitrated Ta components, creating what seems to be an apparent peak shift in \figref{fig:XPSN1}(b). 
This process is a combination of multiple physical and geometrical effects; first, the structure is layered, and the XPS probe depth, while small, is finite and reaches below the surface by several nanometers, revealing that TaN is visible beneath the TiN surface \cite{hofmannS&IASputterDepthProfiling2014,shard2024practical,tougaard2005xps}.
Second, due to the destructive nature of Ar sputtering, it is not possible to easily identify whether or not the contributions from the other Ta phases are the result of a TaN-Ta gradient at the boundary or whether they arise from an Ar sputtering-induced sputter-reduction, a common physical effect in sputtering experiments \cite{shard2024practical,hofmann2014sputter,simpson2017xps}. 
As can be seen from \figref{fig:XPSN1}(e) and \figref{fig:XPSN2}(c), there appears to be a small amount of nitrogen left over, this is due to both the presence of the TaN intermediate layer, but also the preferential sputtering of Ti over N in this energy range \cite{ranjan2001absolute}.

This TaN intermediate layer is supported by the deconvolution of the N 1s and Ti 2p spectra which are shown in \figref{fig:XPSN1}(c) and \figref{fig:XPSN1}(d) respectively; these show the appearance of the TaN peak at higher binding energy (402.3 eV for N 1s and 463.6 eV for Ti 2p) first, then the appearance of the lower binding energy $\mathrm{Ta^0}$ peak (400.9 eV for N 1s and 462.2 eV for Ti 2p). 
In the Ti 2p dataset from \figref{fig:XPSN1}(d), we find that while the sputtering does result in an increase in the intensity of the background as the sputtering progresses, which can possibly be attributed to amorphization, the decomposition from the peak model suggests that the relative intensity of the TiN, TiO$_x$N$_y$, and TiO$_2$ phases remain approximately constant during the sputtering process until the TaN and Ta layers begin to be approached, suggesting that during sputtering oxygen quickly re-adsorbs to the surface while the Ti and N are sputtered away until the TaN interface is revealed.
This is consistent with quantification results using the O 1s region which suggests the continued presence of O until the Ti is completely sputtered away. 
The idea that there is a ‘TiON’-like phase is consistent with the constant presence of a ‘NO’ species in \figref{fig:XPSN1}(c). 
The Ti 2p and N 1s spectral changes after Ar sputtering are therefore consistent with previous studies on Ar-etched and nitrogenated Ti surfaces and provide a robust model for demonstrating the appearance of a TaN interface between the TiN and Ta layers \cite{dieboldSSSTiO2XPS1996,yang2000investigation,bertotiASSSurfaceCharacterisationPlasmanitrided1995}

\begin{figure*}[!htb]
\centering
\includegraphics[width=\textwidth]{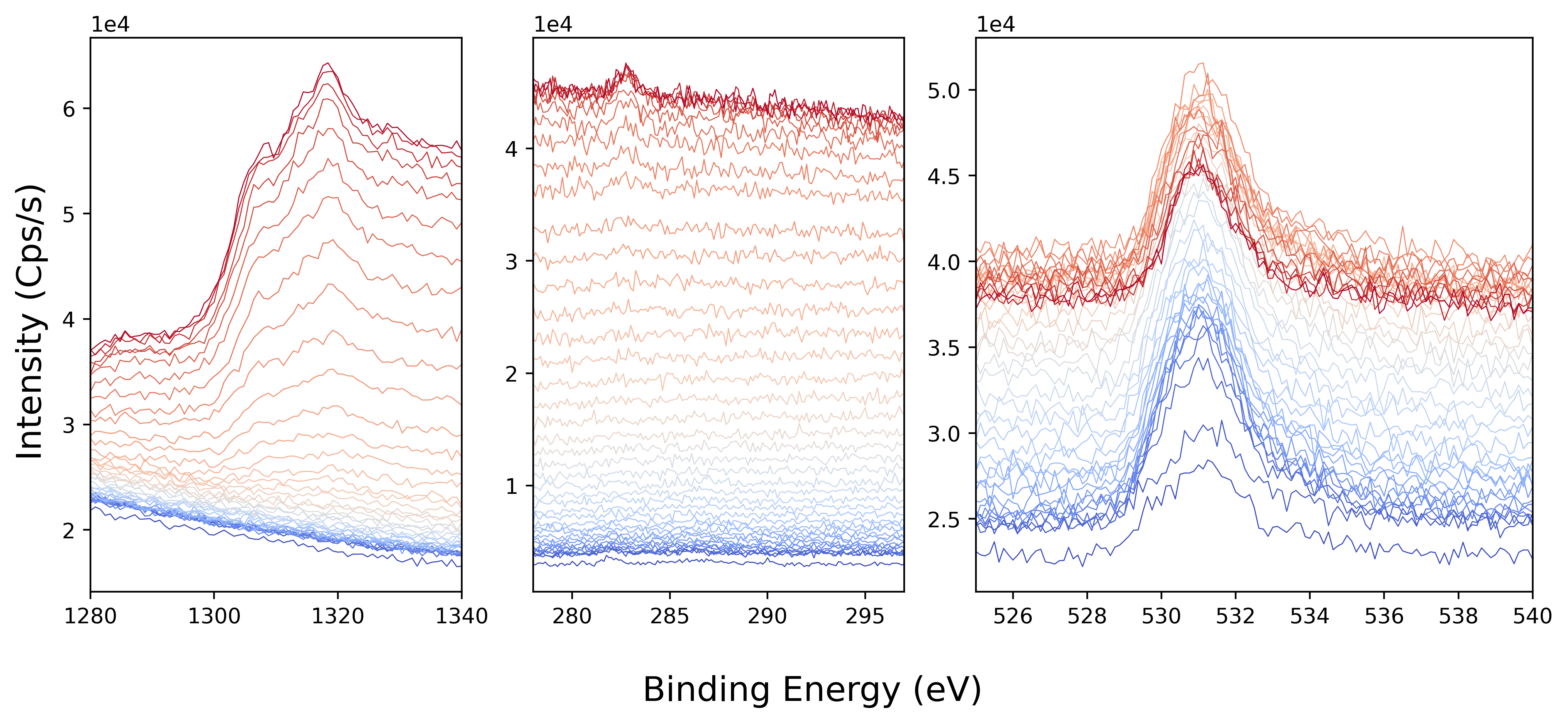}
\caption{(Left) Ta MNN, (Middle) C 1s, and (Right) O 1s narrow scan spectra accumulated during measurement.  Blue corresponds to initial sputtering step and red as the final step; consistent with Fig. \ref{fig:XPSN1}.}
\label{fig:XPSN3}
\end{figure*} 

\subsection{\label{sec:XPSmethod} Data acquisition and sputtering}

X-ray photoelectron spectra were collected, and in situ Argon sputtering was performed, using an ULVAC-PHI Genesis XPS system with a monochromatized Al K$_\alpha$ source (1486.6 eV). 
The vacuum level was maintained at better than $3.2 \times 10^{-7}$ Pa during measurement, which rose intermittently to $1 \times 10^{-6}$ Pa during sputtering steps. 
The X-ray beam spot size was 200 microns in diameter, and the nominal beam power was 50 W at 15 kV, as set by the manufacturer. 
The monoatomic Argon beam was set to 500 eV kinetic energy, with a raster region of $3 \times 3$ $ \mathrm{mm}^2$. 
The Argon beam spot size was 0.5 mm in diameter, with a beam-spot current of 6 nA as set by the manufacturer. The angle of the Argon gun with respect to the sample normal was 45 °.

For the XPS measurements, the integration time per step was 100 ms for all spectra. The pass energy was set to 55 eV for the Ti 2p, N 1s, Ta 4f and Valance band measurements. O 1s and C 1s core levels were also monitored during sputtering with a pass energy of 140 eV to evaluate the presence of trace amounts of adventitious carbon (adC) or the presence of oxidized species or bonds (such as $\mathrm{TiO_x}$, $\mathrm{TaO_x}$, NO, etc.). 
Notably, the pre-sputter Ti 2p spectrum reveals the presence of a thin layer of carbonized Ti, such as TiC or TiCN, which is immediately removed following the first sputter steps and as such were not included in the data set. 
It was found thereafter that there were negligible amounts of C present.
Ta MNN Auger spectra were also acquired with a pass energy of 224 eV, in order to provide a qualitative indication of when the Ta was present at the film surface, due to the increased surface sensitivity of this low kinetic energy peak. 
The C 1s, O 1s, and Ta MNN Auger spectra are available in Figure \ref{fig:XPSN3}.

The sputtering procedure consisted of an initial ‘rough sputtering’ with steps of 10 min for 5 steps, following by ‘fine sputtering’ steps of 1 min for 162 steps.
This was done since the initial TiN film was expected to be 40 nm thick, and it was previously found using survey spectra (Pass energy 112 eV, same XPS settings as previous) that after 80 min of sputtering, a weak Ta signal became visible. 
The effective sputter rate was determined by comparing the pitch of the crater created on the 40 nm TiN film, which we estimate to be slightly less than 0.25 nm per min. 

\subsection{\label{sec:XPSndataalysis} Data postprocessing}

Corrected empirical relative sensitivity factors (ce-RSFs) were obtained from the Multipak software on the instrument and used without further correction.
Analysis of the acquired spectra was performed in Python using the Lmfit and LmfitXPS packages \cite{newville2016lmfit,hochhaus_2025_15130863}. 
A static Tougaard background was subtracted from the N 1s and Ta 4f peaks used for quantification \cite{tougaard2021practical,major2020practical,sherwood2019use}.
This background was determined from the data as filtered by a median filter with a kernel size of 7 points, before being subtracted from the unfiltered data. 
An active Tougaard background was used for the Ti 2p spectra without the use of a median filter. 

Valence band XPS measurements, presented in \ref{fig:XPSN2} were used to provide further information on the metallicity of the sample and binding energy calibration to the fermi level. 
Additionally, valence band XPS also provides a fingerprint of the material phases present \cite{edaS&IAXrayPhotoelectronSpectroscopybased2022,engelhardIntroductoryGuideBackgrounds2020,baerJVSTAPracticalGuidesXRay2019}. 

For the Ta 4f, a set of Doniach-Sunjic doublets was used to model the different possible Tantalum nitride phases, including ``hexagonal or cubic'' TaN, ``substoichiometric'' $\mathrm{Ta_2N}$, ``doped'' $\mathrm{TaN_{0.05}}$, and Ta0, with a small contribution from oxidized tantalum $\mathrm{TaO_x}$. This was taken from references \cite{arranz2000tantalum,arranz2005composition}.
The asymmetry parameter was set to 0.05, the spin-orbit coupling was constrained to be 1.9 eV, the relative area of the Ta $\mathrm{4f_{7/2}}$ and Ta $\mathrm{4f_{5/2}}$ peaks was set to the theoretical value of 3/4, and the Coster-Kronig parameter was set to 1. The additional parameters are in \tabref{tab:XPST1}. 

For quantitative analysis, the N 1s and Ti 2p spectral decomposition also includes a voigt for the Ta $\mathrm{4p_{3/2}}$ and Ta $\mathrm{4p_{1/2}}$ core levels. 
The analysis is complicated by the presence of the Ta 4p core levels, which have considerable peak overlap between the Ta 4p states and the tails of the Ti 2p and N 1s spectra. 
Therefore, the curve-fitting of the Ti 2p and N 1s was performed together with these Ta core levels, and only the area assigned as originating from the Ti 2p and N 1s core levels was used in the quantification \cite{shard2020practical}.

\begin{table*}[htb!] 
\renewcommand{\arraystretch}{1.3}
    \setlength{\tabcolsep}{4pt}
    \centering
    \begin{tabular}{|c|c|c|c|c|c|c|}
\hline 
 Component                & Binding Energy & Lorentzian $\sigma$        & Gaussian $\sigma$        & FWHM      & Lineshape Function      & Constraints   \\
\hline 
 Ta 4f                    &          &                  &                 &          &                         &       \\
\hline Ta$^0$                      & 22.1     & 0.08              & 0.38            & 0.76      & D.S. Dublett            &               \\
\hline $\beta$-TaN$_{0.05}$     & 22.7     & 0.13              & 0.42            & 0.93      & D.S. Dublett            &               \\
\hline $\gamma$-Ta$_2$N         & 23.3     & 0.13              & 0.42            & 0.92      & D.S. Dublett            &               \\
\hline Cubic TaN                & 24.0     & 0.25              & 0.55            & 1.40      & D.S. Dublett            &               \\
\hline TaO$_x$                  & 26.8     & 0.25              & 0.38            & 1.15      & D.S. Dublett            &               \\
\hline Ti 2p                    &          &                   &                 &           &                         &               \\
\hline TiN                      &          &                   &                 &           & Composite of:           & Relative Area \\
\hline + Ti $2p_{3/2}$          & 455*     & 1.18              & 0.66            & 1.3       & Pseudo-Voigt  GLP(30)   & 5             \\
\hline + Shake ($2p_{3/2}$)     & 457.2    & 3.47              & 1.93            & 3.8       & Pseudo-Voigt  GLP(30)   & 11            \\
\hline + Ti $2p_{1/2}$          & 460.9    & 1.18              & 0.66            & 1.3       & Pseudo-Voigt    GLP(30) & 2.6           \\
\hline + Shake ($2p_{1/2}$)     & 462.9    & 3.47              & 1.93            & 3.8       & Pseudo-Voigt    GLP(30) & 5.7           \\
\hline + Plasmon ($2p_{3/2}$)   & 467.5    & 6.66              & 3.7             & 7.3       & Pseudo-Voigt    GLP(30) & 3.5           \\
\hline + Plasmon ($2p_{1/2}$)   & 474.0    & 6.66              & 3.7             & 7.3       & Pseudo-Voigt    GLP(30) & 1.7           \\
\hline TiO$_x$N$_y$             &          &                   &                 &           &                         &               \\
\hline + $2p_{3/2}$             & 455.8    & 1.18              & 0.66            & 1.3       & Pseudo-Voigt    GLP(30) & 2             \\
\hline + $2p_{1/2}$             & 461.5    & 2.20              & 1.21            & 2.4       & Pseudo-Voigt    GLP(30) & 1             \\
\hline TiO$_2$                  &          &                   &                 &           &                         &               \\
\hline + $2p_{3/2}$             & 459.2    & 1.18              & 0.66            & 1.3       & Pseudo-Voigt    GLP(30) & 2             \\
\hline + $2p_{1/2}$             & 465.8    & 2.20              & 1.21            & 2.4       & Pseudo-Voigt    GLP(30) & 1             \\
\hline Ta$^0$ $4f_{5/2}$        & 462.2    & 2.16              & 1               & 7.00      & Voigt                   &               \\
\hline TaN $4f_{5/2}$           & 463.6    & 2.16              & 1               & 7.00      & Voigt                   &               \\ 
\hline N 1s                     &          &                   &                 &           &                         &               \\
\hline TiN                      & 397.12   & 0.2               & 0.42            & 0.74      & Voigt                   & 1             \\
\hline + Plasmon 1              & 404.75   & 0.15              & 0.73            & 0.80      & Voigt                   & 0.0175        \\
\hline + Plasmon 2              & 408.5    & 0.15              & 1.45            & 1.26      & Voigt                   & 0.0175        \\
\hline TaN                      & 397.9    & 0.2               & 0.42            & 0.74      & Voigt                   &               \\
\hline NO                       & 399.6    & 0.7               & 0.7             & 2.23      & Voigt                   &               \\
\hline Ta0 $4f_{7/2}$           & 400.9    & 2.08              & 0.7             & 6.3       & Voigt                   &               \\
\hline TaN $4f_{7/2}$           & 402.3    & 2.08              & 0.7             & 6.3       & Voigt                       &           \\
\hline
    \end{tabular}
\caption{Fit parameters for the lineshapes used in quantification of the TiN/TaN/Ta interface as a function of sputtering time. All values are in units of eV. A '+' symbol indicates the peak is a component of the lineshape above it.}
\label{tab:XPST1}
\end{table*}

For the Ti 2p spectra a model composed of TiN, defective sub-stoichiometric surface TiO$_x$N$_y$, TiO$_2$, Ta$^0$, and TaN was used, with a reference lineshape for TiN, which accounts for its multiplet structure, found from the literature \cite{jaeger2012complete,bertotiASSSurfaceCharacterisationPlasmanitrided1995}.
The oxidized Ti species, TiO$_x$N$_y$ and TiO$_2$, were expected as Ti is a well-known ‘getter’ and highly reactive with oxygen; each of these components were accounted by a doublet of product pseudo-Voigt  functions with spin-orbit splitting, binding energies, and full-width-half-maxima (FWHM) from the literature \cite{bertoti1995surface,aliJMSEffectContentPhase2023,de2005multiplet,bagus2016multiplet,shard2024practical,soundiraraju2017two,jena2023investigation,carleyJCSFT1IdentificationCharacterisationMixed1987,dieboldSSSTiO2XPS1996}. 
The Ta $\mathrm{4p_{1/2}}$ contributions were accounted for by including two Voigt functions corresponding to the Ta0 and TaN phases.
For simplicity, we restrict our model to two Ta 4p pseudo-voigt functions, with the Ta $\mathrm{4p_{1/2}}$ peaks set to 462.2 and 463.6 eV for $\mathrm{Ta}^0$ and TaN, respectively.
While in principle a distinct Voigt function for each chemically distinct Ta phase should be used, in practice the chemical shifts corresponding to the intermediate phases are difficult to determine because the Ta 4p orbitals are highly screened and have large FWHM. 

For the N 1s spectra, multiple components were used to model the TiN-nitrogen peak. The TiN-N 1s peak included one main peak at 397.1 eV and two plasmons at 404.75 and 408.5 eV, and these were constrained to have an amplitude of 1.75\% of the main peak \cite{jaeger2012complete}. 
The TaN-based nitrogen peak was assigned to 397.9 eV, and a nitrogen-oxygen contaminant peak was identified at 399.6 eV; these binding energies were taken from references in the literature \cite{arranz2000tantalum,arranz2005composition}.
The Ta $\mathrm{4p_{3/2}}$ peaks were set to 400.9 and 402.3 eV for $\mathrm{Ta}^0$ and TaN, respectively, consistent with the Ti 2p spectra analysis.

The components for each XPS core level data are summarized in \tabref{tab:XPST1}. Quantification, shown in \ref{fig:XPSN1}(e), was done according to the standard procedure after background subtraction without QUASES-type correction \cite{tougaard1986quantitative,tougaard2005xps,baer2025more,hesse2015improved,hajati2008nondestructive}.

\FloatBarrier 

\bibliography{apssamp,
BibReferenceFiles/Modular_QuantumCompLit.bib,
BibReferenceFiles/XPS_ProfileLit,
BibReferenceFiles/Intro}

\end{document}